\documentclass[aps,prx,preprint,onecolumn,citeautoscript,footinbib,
eqsecnum]{revtex4-1}
\usepackage{graphicx}
\usepackage{color}
\usepackage[dvipsnames]{xcolor}
\usepackage[papersize={8.5in,11in}]{geometry}
\usepackage[colorlinks=true]{hyperref}
\usepackage{ulem}
\usepackage{subfig}
\usepackage{tikz-feynman}
\usepackage[mathscr]{euscript}
\usepackage[section]{placeins}
\usepackage{comment}
\usepackage{bigints}
\usepackage[font=footnotesize,justification=raggedright,singlelinecheck=false]{caption}
\hypersetup{
    bookmarks=true,         
    unicode=false,          
    pdftoolbar=true,        
    pdfmenubar=true,        
    pdffitwindow=false,     
    pdfstartview={FitH},    
    pdfsubject={},   
    pdfcreator={},   
    pdfproducer={}, 
    pdfkeywords={} {} {}, 
    pdfnewwindow=true,      
    colorlinks=true,       
    linkcolor=magenta, 
    citecolor=blue,        
    filecolor=magenta,      
    urlcolor=blue           
}

\usepackage{latexsym}
\usepackage{bbm}
\usepackage[export]{adjustbox}
\usepackage[mathscr]{euscript}
\usepackage{amsfonts,amsthm,amsmath,amssymb,upgreek}
\usepackage{amsmath,amssymb,bm,bbm}
\newcommand{\beq}{\begin{equation}}
\newcommand{\eeq}{\end{equation}}
\newcommand{\ba}{\begin{array}{ccc}}
\newcommand{\ea}{\end{array}}

\newcommand{\calE}{\mathscr{E}}

\def\bea{\begin{eqnarray}}
\def\eea{\end{eqnarray}}
\usepackage[colorlinks=true]{hyperref}
\hypersetup{
    bookmarks=true,         
    unicode=false,          
    pdftoolbar=true,        
    pdfmenubar=true,        
    pdffitwindow=false,     
    pdfstartview={FitH},    
    pdftitle={My title},    
    pdfauthor={Author},     
    pdfsubject={Subject},   
    pdfcreator={Creator},   
    pdfproducer={Producer}, 
    pdfkeywords={keyword1} {key2} {key3}, 
    pdfnewwindow=true,      
    colorlinks=true,       
    linkcolor=magenta, 
    citecolor=blue,        
    filecolor=magenta,      
    urlcolor=cyan           
}

\def\Xint#1{\mathchoice
   {\XXint\displaystyle\textstyle{#1}}%
   {\XXint\textstyle\scriptstyle{#1}}%
   {\XXint\scriptstyle\scriptscriptstyle{#1}}%
   {\XXint\scriptscriptstyle\scriptscriptstyle{#1}}%
   \!\int}
\def\XXint#1#2#3{{\setbox0=\hbox{$#1{#2#3}{\int}$}
     \vcenter{\hbox{$#2#3$}}\kern-.5\wd0}}

\def\dashint{\Xint-}

\newcommand{\ff}{\mathfrak{f}}
\newcommand{\fb}{\mathfrak{b}}

\renewcommand{\approx}{\simeq}
\renewcommand{\Re}{\text{Re}}
\renewcommand{\Im}{\text{Im}}

\newcommand{\ce}{\calE}
\newcommand{\rd}{{\rm d}}
\newcommand{\sgn}{{\rm sgn\,}}
\newcommand{\Tr}{{\rm \, Tr\,}}

\makeatletter
\newcommand{\mathleft}{\@fleqntrue\@mathmargin0pt}
\newcommand{\mathcenter}{\@fleqnfalse}
\makeatother

\tikzset{
  mid arrow/.style={postaction={decorate,decoration={
        markings,
        mark=at position .575 with {\arrow[#1]{stealth}}
      }}},
  near arrow/.style={postaction={decorate,decoration={
        markings,
        mark=at position .275 with {\arrow[#1]{stealth}}
      }}},
   far arrow/.style={postaction={decorate,decoration={
        markings,
        mark=at position .800 with {\arrow[#1]{stealth}}
      }}},
}

\newcommand*\pFqskip{8mu}
\catcode`,\active
\newcommand*\pFq{\begingroup
        \catcode`\,\active
        \def ,{\mskip\pFqskip\relax}%
        \dopFq
}
\catcode`\,12
\def\dopFq#1#2#3#4#5{%
        {}_{#1}\textbf{F}_{#2}\biggl(\genfrac..{0pt}{}{#3}{#4};#5\biggr)%
        \endgroup
}

\begin{document}
\title{Excitation spectra of quantum matter\\ without quasiparticles II: random $t$-$J$ models}

\author{Maria Tikhanovskaya, Haoyu Guo,  Subir Sachdev, Grigory Tarnopolsky}
\affiliation{Department of Physics, Harvard University, Cambridge MA, 02138, USA}

\date{\today}

\begin{abstract}
We present numerical solutions of the spectral functions of $t$-$J$ models with random and all-to-all exchange and global SU($M$) spin rotation symmetry. The solutions are obtained from the saddle-point equations of the large volume limit, followed by the large $M$ limit. These saddle point equations involve Green's functions for fractionalized spinons and holons carrying emergent U(1) gauge charges, obeying relations similar to those of the Sachdev-Ye-Kitaev (SYK) models. The low frequency spectral functions are compared with an analytic analysis of the operator scaling dimensions, with good agreement. We also compute the low frequency and temperature behavior of gauge-invariant observables: the electron Green's function, the local spin susceptibility and the optical conductivity; along with the temperature dependence of the d.c. resistivity. The time reparameterization soft mode (equivalent to the boundary graviton in holographically dual models of two-dimensional quantum gravity) makes important contributions to all observables, and provides a linear-in-temperature contribution to the d.c. resistivity.

\end{abstract}
\maketitle

\tableofcontents

\section{Introduction}
\label{sec:intro}

This paper extends our recent analysis in Ref.~\cite{Tikhanovskaya:2020elb} from insulating spin models to metallic states of a random $t$-$J$ model. This yields a rare solvable model of a metallic system without quasiparticle excitations, and so it is worthwhile to explore its properties in detail.

In the condensed matter literature, extensions of the Sachdev-Ye-Kitaev (SYK) models \cite{SY92,kitaev2015talk,SS15} provide solvable theories of `strange' and `bad' metal behavior. Extensions which include single particle
hopping \cite{Balents2017,Zhang2017,Chowdhury2018,Patel2017,Patel2019} have regimes of intermediate temperature ($T$) with the paradigmatic linear-in-$T$ resistivity, but with a bad metal resistivity {\it i.e.} larger than the quantum of resistance $h/e^2$ (in two dimensions).
Upon lowering $T$, these models enter a disordered Fermi liquid regime with resistivity $ \sim T^2$ and well-defined quasiparticles, once the resistivity becomes smaller than $h/e^2$. A strange metal regime, {\it i.e.\/} a linear-in-$T$ resistivity smaller than $h/e^2$, only appears possible in special models with `resonant' interactions \cite{Patel2019}.

In recent work, Joshi {\it et al.\/} \cite{Joshi:2019csz} argued that a shortcoming of the above SYK models \cite{Balents2017,Zhang2017,Chowdhury2018,Patel2017,Patel2019} is the absence of `Mottness' {\it i.e.\/} a strong local repulsion between electrons that allows an interaction-driven insulator at commensurate densities.
Mottness was present in the original Sachdev-Ye model \cite{SY92} of an insulator, and this can be extended to metals by considering a SYK-like version of the $t$-$J$ model with random exchange interactions \cite{PG98,Joshi:2019csz,Guo:2020aog,Shackleton20}. Ref.~\cite{Joshi:2019csz} argued that upon tuning the electron density, such $t$-$J$ models have critical points or phases in which the SYK criticality extends down to $T=0$.

Ref.~\cite{Joshi:2019csz} also proposed a large $M$ limit of a $t$-$J$ model with SU($M$) symmetry, when the solution reduces to a set of SYK-like saddle-point equations for the Green's functions of {\it fractionalized\/} particles carrying emergent U(1) gauge charges: these are the spinons and holons carrying the spin and charge of the underlying electrons. The leading low frequency behavior of the Green's functions were determined in Ref.~\cite{Joshi:2019csz}. Here, we shall provide a full numerical analysis of the saddle-point equations, and show that solutions with the proposed low frequency singularities do exist over some range of parameters. We shall also analytically compute the subleading corrections to the low frequency behavior, along the lines of Refs.~\cite{Guo:2020aog,Tikhanovskaya:2020elb}, and show that it is consistent with our numerical results.

With the knowledge of the Green's functions of the fractionalized particles, we can compute various physical gauge-invariant observables. We list the properties of the electron spectral weight, $\rho_{e} (\omega)$, the spin spectral density $\rho_Q(\omega)$, and the optical conductivity $\sigma(\omega)$ at frequency $\omega$ and at temperature $T=0$ ($J$ is the root-mean-square exchange interaction):
\bea
 \rho_{e} (\omega) &=& A^\pm\left(1-\sum_h A_h^\pm |\omega/J|^{h-1}-\sum_{hh'}A_{hh'}^{\pm}|\omega/J|^{h+h'-2}\right)\,,  \nonumber \\
\rho_Q(\omega) &=& B^{\pm}\left(1-\sum_h B_h |\omega/J|^{h-1}-\sum_{hh'}B_{hh'}|\omega/J|^{h+h'-2}\right)\,,\nonumber \\
\mbox{Re}\, \sigma (\omega) &=& \sigma_0\left(1-\sum_h C_h |\omega/J|^{h-1}-\sum_{hh'}C_{hh'}|\omega/J|^{h+h'-2}\right)\,, \label{e1}
\eea
where $\pm$ superscripts refer to $\omega \gtrless 0$.
  Here the summation is over the spectrum of irrelevant operators, and $A,B,C,\sigma_0$'s are coefficients that will be given in the main text in \eqref{eq:rhocT=0}, \eqref{eq:rhoQT=0} and \eqref{eq:sigmaT=0}.
In the above expressions, we include the leading term arising from the conformal solution, and the subleading corrections. Example of the latter includes a leading irrelevant operator with scaling dimension $h_1 < 2$ (see Fig.~\ref{fig:hsplot}), and the time reparameterization mode ({\it i.e.\/} the boundary graviton in dual models of two-dimensional quantum gravity \cite{Maldacena:2016hyu,Maldacena:2016upp,kitaev2017}) with scaling dimension $h_0=2$. We note that the presence of the mode with $h_1 < 2$ is a new feature of the doped $t$-$J$ model, and was not present in the undoped antiferromagnet examined earlier \cite{Tikhanovskaya:2020elb}; the leading irrelevant operator in the latter case was the $h_0=2$ boundary graviton.

\begin{figure}
  \centering
  \includegraphics[width=\textwidth]{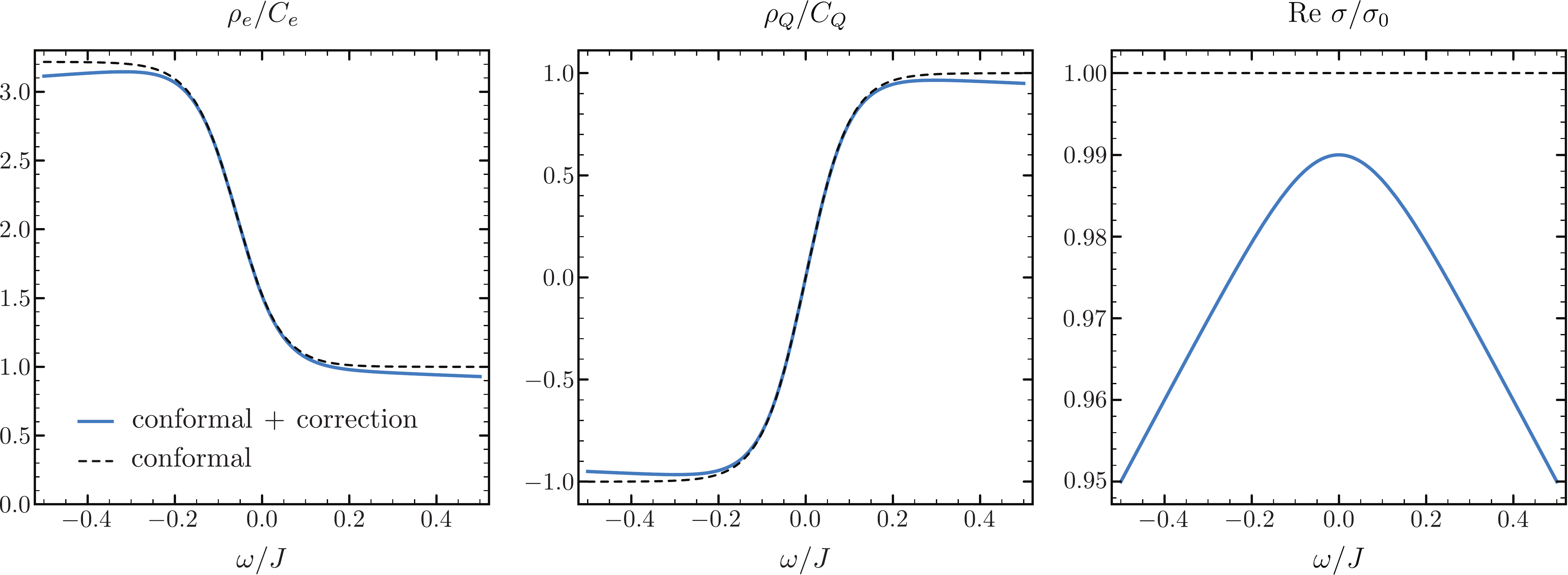}
  \caption{Plots of the functions in Eq.~\eqref{e2}. We used $\beta J=20$, $\theta_f=0.1\pi$, $\theta_b=0.3\pi$, $\alpha_{0}=0.05$.}\label{fig:eq1.2}
\end{figure}

We can also extend the results in (\ref{e1}) to non-zero $T$: each term will be multiplied by a scaling function of $\omega/T$. The expressions for these scaling functions are rather complicated at general scaling dimension $h$ and are given in the body of the paper. But the scaling functions do simplify for the $h_0=2$ case. Including only these $T$-dependent corrections, the non-zero $T$ form of (\ref{e1}) is, for $\omega, T \ll J$ ($\beta \equiv 1/T$)
\begin{align}
&\rho_{e} (\omega) = \frac{C_{e} \cosh(\frac{\beta\omega}{2})}{J\cosh(\frac{\beta\omega}{2}-\pi\calE_{e})}
\left(1-\frac{\pi \alpha_{0}}{\beta J}\Big(\frac{\beta\omega}{2\pi}-\calE_{e}\Big)\big(4\tanh(\frac{\beta\omega}{2}-\pi\calE_{e}) -3(\sin(2\theta_f)-\sin(2\theta_b))\big)\right)\,,\nonumber \\
&\rho_Q (\omega) = \frac{C_Q}{J} \tanh\big(\frac{\beta \omega}{2}\big)\left( 1 -\frac{2\alpha_{0}\omega}{J} \tanh\big(\frac{\beta \omega}{2}\big)\right)\,, \nonumber \\
&\mbox{Re}\, \sigma (\omega) = \sigma_0\left(1-\frac{2\alpha_{0} \omega}{J}\coth(\frac{\beta\omega}{2})\right)\,.\label{e2}
\end{align}
In these expressions, the terms outside the square brackets arise from the leading conformal theory, while the terms inside the square brackets are the contributions of the boundary graviton mode (with $h_0 = 2$). Here $C_{e}$, $C_Q$ and $\sigma_0$ are normalization constants that depend on the UV details. $\alpha_{0}$ is a dimensionless number parameterizing the strength of $h_0=2$ correction: this is also a UV parameter and is extracted from the numerics.  $\theta_f$ and $\theta_b$ parameterize the conformal solution \eqref{eq:gft=0} and are related to the U(1) charges. $\ce_{e}$ is the spectral asymmetry of the electron Green's function \eqref{eq:get=0}, which can be calculated from $\theta_f$ and $\theta_b$. We also include  schematic plots of the functions in \eqref{e2} in Fig.~\ref{fig:eq1.2}.

Finally, we present the results for the d.c. resistivity, $\rho(T)$. Note from (\ref{e1}) that the zero frequency conductivity from the scaling limit solution is a constant {\it i.e.} the conformal solution yields a residual resistivity. So all the $T$ dependence arises from corrections to conformality, and these are (from (\ref{eq:sigmaDCT}))
\begin{equation}
\rho(T) =\frac{1}{\sigma_0}\left[1+\sum_{h}\alpha_h(v_{hf+}+v_{hf-}+v_{hb+}+v_{hb-}) R(h) \left(\frac{T}{J}\right)^{h-1} \right]\,.
\label{erho}
\end{equation}
Note that the time reparameterization mode, with $h_0=2$, yields a linear-in-temperature correction. Further results on the numerical values of the coefficients in (\ref{e1}) and (\ref{erho}) appear in Section~\ref{sec:Numerics}.

The outline of our paper is as follows. We define the $t$-$J$ model in Section~\ref{sec:tJmodel}, along with its large $M$ limit using fractionalized spinons and holons. We also present the SYK-like saddle-point equations and their conformal solutions. Section~\ref{sec:OperSpectrum} turns to our new results on the operator spectrum and the associated corrections to conformality for the spinon and holon Green's functions. We also find regimes where the conformal solutions are unstable: this can happen by the appearance of operators with scaling dimension $h<1$, and by complex values of $h$. Section~\ref{sec:GIO} combines the fractionalized results of Section~\ref{sec:OperSpectrum} to compute the spectral functions of various gauge invariant observables and their corrections to conformality. Finally, Section~\ref{sec:Numerics} computes the full numerical solution of the saddle-point equations and compares them with our conformal expansions.

\section{$t$-$J$ model and fractionalization}
\label{sec:tJmodel}

\subsection{Model Hamiltonian}
We consider the following Hamiltonian of the $t$-$J$ model
\begin{equation}\label{}
  H=\sum_{\left\langle ij\right\rangle,l,\alpha}t_{ij}c_{il\alpha}^{\dagger}c_{jl\alpha}+h.c.+\sum_{\left\langle ij\right\rangle,\alpha\beta}J_{ij}\left(S_{i\alpha\beta}S_{j\beta\alpha}-\frac{1}{M}S_{i\alpha\alpha}S_{j\beta\beta}\right).
\end{equation}Here $c_{il\alpha}$ is electron operator, with $i$ denoting site, $l$ denotes an auxiliary SU$(M')$ orbital label, and $\alpha$ denotes SU($M$) spin. The $S_{i\alpha\beta}=S_{i\beta\alpha}^{\dagger}$ is the spin operator on site $i$, and the $1/M$ term (which will be dropped in large $M$ limit) is added to ensure it transform in the adjoint of SU$(M)$.

We assume the system lives on a Bethe lattice of coordination number $z$ with
\begin{equation}\label{}
  t_{ij}=\frac{t}{\sqrt{Mz}}\,,
\end{equation}
and $J_{ij}$'s are Gaussian random variables with
\begin{equation}\label{}
  \overline{J_{ij}}=0, \quad \overline{J_{ij}^{2}}=\frac{J^2}{M z}\,.
\end{equation}

Implementing dynamical mean-field approximation ($z\to\infty$), and assuming a diagonal mean-field ansatz, the Hamiltonian induces the following interaction terms in the single-site action
\begin{equation}\label{}
\begin{split}
  S\supset& \frac{t^2}{M}\sum_{l \alpha}\int_{0}^{1/T}\rd \tau\rd \tau'\sum_{l\alpha}c_{l\alpha}^{\dagger}G_c(\tau,\tau')c_{l\alpha}(\tau') - \frac{J^2}{2M} \sum_{\alpha\beta}\int_{0}^{1/T}\rd \tau\rd \tau' S_{\alpha\beta}(\tau)Q(\tau,\tau')S_{\beta\alpha}(\tau')\,.
\end{split}
\end{equation}

Here the mean-field ansatz $G_{e}$ and $Q$ should satisfy the following self-consistency equations
\begin{equation}
  G_{e}(\tau,\tau') = \frac{-1}{M M'}\sum_{l\alpha}\langle c_{l\alpha}(\tau)c_{l\alpha}^{\dagger}(\tau')\rangle \,, \quad
  Q(\tau,\tau') = \frac{1}{M^2}\sum_{\alpha\beta}\left\langle S_{\beta\alpha}(\tau)S_{\alpha\beta}(\tau')\right\rangle \,.
\end{equation}
We also assume large $M,M'$ limit with fixed
\begin{equation}\label{}
  k=\frac{M'}{M}\,.
\end{equation}
The full action also contains WZW terms for slave particles and Langrangian multipliers, which will be detailed shortly.

\subsection{Fermionic spinon + Bosonic holon}
\label{sec:fermionicspinons}

We fractionalize the model by introducing fermionic spinon and bosonic holon:
\begin{equation}
   c_{il\alpha} =f_{i\alpha}b_{il}^{\dagger}, \quad
  S_{i\alpha\beta} = f_{i\alpha}^{\dagger}f_{i\beta}\,.
\end{equation}
Here the boson $b_{il}$ transforms in anti-fundamental of SU($M'$) such that the electron $c_{il\alpha}$ transforms in fundamental of SU($M'$). This induces a U(1) gauge symmetry
\begin{equation}\label{}
  f_{i\alpha}(\tau)\to f_{i\alpha}(\tau)e^{i \phi_i(\tau)},\quad b_{il}(\tau)\to b_{il}(\tau)e^{i\phi_i(\tau)}.
\end{equation}
For the theory to be consistent, the physical Hilbert space must be U(1) gauge-symmetric, which implies that the gauge charge is conserved, and we consider the representation
\begin{equation}\label{}
  \sum_{\alpha}f_{i\alpha}^\dagger f_{i\alpha}+\sum_{l}b_{il}^{\dagger} b_{il}=\kappa M.
\end{equation}
We also implement the doping density $p$ by
\begin{equation}\label{}
  \frac{1}{M'}\sum_{l} \langle b_{il}^\dagger b_{il}\rangle=p.
\end{equation}

The full action is thus
\begin{eqnarray}
  S[f,b,\lambda] &=& \int_0^{1/T} d \tau \left[ \sum_{\ell} b_{\ell}^\dagger \left( \frac{\partial}{\partial \tau} + i \lambda \right) b_{\ell}
+ \sum_{\alpha} f_{\alpha}^\dagger \left( \frac{\partial}{\partial \tau} + \epsilon_0 + i \lambda \right) f_{\alpha} - i \lambda \frac{M}{2} \right] \nonumber \\
&~&~+\frac{t^2}{M} \sum_{\ell,\alpha} \int_{0}^{1/T} d \tau d \tau'
f^\dagger_{\alpha}(\tau)b_{\ell}(\tau) G_c(\tau-\tau') b^\dagger_{\ell}(\tau') f_{\alpha}(\tau')\nonumber
\\ &~&~-
\frac{J^2}{2M} \sum_{\alpha,\beta} \int_{0}^{1/T} d \tau d \tau'  Q(\tau-\tau') f_{\alpha}^\dagger (\tau) f_{\beta} (\tau) f_{\beta}^\dagger (\tau') f_{\alpha} (\tau') \,.
\end{eqnarray}
Introducing bilocal fields
\begin{equation}
  G_f(\tau,\tau') = \frac{-1}{M}\sum_{\alpha}f_{\alpha}(\tau) f^\dagger_{\alpha}(\tau'), \quad
  G_b(\tau,\tau') = \frac{-1}{M'}\sum_{l}b_l(\tau)b^\dagger_l(\tau')
\end{equation}
the saddle-point equations for the slave particles $b_{\ell}$ and $f_\alpha$ are given as follows,
\begin{eqnarray}
  G_b(i\omega_n) &=& \frac{1}{i\omega_n+\mu_b-\Sigma_b(i\omega_n)}, \label{Eq:EoM1} \\
  \Sigma_b(\tau) &=& -t^2G_f(\tau)G_f(-\tau)G_b(\tau) \label{Eq:EoM2},\\
  G_f(i\omega_n) &=& \frac{1}{i\omega_n+\mu_f-\Sigma_f(i\omega_n)} \label{Eq:EoM3},\\
  \Sigma_f(\tau) &=& -J^2 G_f(\tau)^2G_f(-\tau)+kt^2G_f(\tau)G_b(\tau)G_b(-\tau) \label{Eq:EoM4}.
\end{eqnarray}
Here $\mu_f$ and $\mu_b$ are chemical potentials, determined by $\epsilon_0$ and the saddle point value of $\lambda$, and chosen to satisfy
\beq
\left\langle f^\dagger f \right\rangle = \kappa-k p \quad, \quad  \left\langle b^\dagger b \right\rangle = p \,. \label{lutt1}
\eeq
The electron Green's functions is a product
 \begin{equation}
     G_{e} (\tau) = - G_f (\tau) G_b (-\tau)\,.
  \label{Eq:EoM5}
 \end{equation}

Eqs.\eqref{Eq:EoM1}-\eqref{Eq:EoM4} admit the following infra-red (IR) conformal saddle point solution at zero temperature \cite{Joshi:2019csz}
\begin{equation}\label{eq:gft=0}
 \quad G^{c}_{a}(\tau) =
-\begin{pmatrix}
e^{\pi \calE_{a} } \\
\zeta e^{-\pi \calE_{a}}
\end{pmatrix} \frac{b_{a}^{\Delta_{a}}}{|J\tau|^{2\Delta_{a}}},\quad G_{a}^{c}(i\omega) = -\frac{i C_{a}}{J}\begin{pmatrix}
e^{-i\theta_{a}}\\
- e^{i\theta_{a}}
\end{pmatrix} |\omega/J|^{2\Delta_{a}-1}\,,
\end{equation}
where $c$ in the superscript means conformal and subscript $a=b,f$ indexes boson and fermion respectively.
We are following the convention of writing functions
as two component vectors, where the first component is for $\tau>0$ or $\omega >0$, and the second one is for $\tau<0$ or $\omega <0$. Also $\calE_{b}$, $\calE_{f}$ and $\theta_{b},\theta_{f}$ are asymmetry parameters and angles and $\zeta$ factor is $\zeta_{b}=1$ and $\zeta_{f}=-1$.  To pass from the coordinate space to the frequency one we use the Fourier transform in our plus/minus basis
\begin{equation}
\bigintssss
\begin{pmatrix}
a_+ \\
a_-
\end{pmatrix} |\tau|^{-\alpha}  e^{i\omega \tau} d\tau =
M(\alpha)\begin{pmatrix}
a_+ \\
a_-
\end{pmatrix} |\omega|^{\alpha-1}\,,
\quad M(\alpha) = \Gamma(1-\alpha) \begin{pmatrix}
i^{1-\alpha} & ~i^{\alpha-1} \\
i^{\alpha-1} & ~i^{1-\alpha}
\end{pmatrix}\,.
\end{equation}

We will restrict our attention here to the the case where the boson and fermion scaling dimensions are $\Delta_f=\Delta_b=1/4$, which is the case proposed for the deconfined critical point in Ref.~\onlinecite{Joshi:2019csz}.
The large $M$ equations also admit a solution in a critical phase \cite{Joshi:2019csz} with $1/4 < \Delta_f < 1/2$, $\Delta_f + \Delta_b = 1/2$ which we will not explicitly consider here: we expect similar results to also apply to this critical phase. For the case $\Delta_f=\Delta_b = 1/4$,
the pre-factors satisfy
\begin{equation}\label{eq:CfCb}
  \begin{split}
     &(t^2/J^{2}) C_f^2 C_b^2\cos(2\theta_f)  =\pi, \\
      & C_f^4 \cos(2\theta_f)-k(t^2/J^{2}) C_f^2C_b^2\cos(2\theta_b)=\pi
  \end{split}
\end{equation}
and we can find
\begin{equation}\label{eq:CfCbsol}
C_{f} = \Big(\frac{\pi (1-\eta)}{\cos(2\theta_{f})}\Big)^{1/4}, \quad C_{b} = \frac{J}{t}\Big(\frac{\pi}{\cos(2\theta_{f})(1-\eta)}\Big)^{1/4}, \quad \eta \equiv -k \frac{\cos (2\theta_{b})}{\cos (2\theta_{f})}\,.
\end{equation}
Positivity constraints on the spectral densities impose restrictions on asymmetry angles: $-\pi\Delta_f<\theta_f<\pi\Delta_f$, $\pi\Delta_b<\theta_b<\pi/2$. Therefore we see that $\eta > 0$ and since
 $C_{b}$ and $C_{f}$ are defined to be real positive numbers, we find another constraint for $\theta_{b}$ and $\theta_{f}$:
\begin{equation}\label{eq:conformal_condition}
   \eta < 1\,.
\end{equation}
For completeness we also list expressions for $b_{f}$ and $b_{b}$:
\begin{equation}
b_{f} = \frac{\cos(2\theta_{f})}{4\pi}(1-\eta), \quad b_{b} = \frac{\cos(2\theta_{f})}{4\pi} \frac{\eta^{2}J^{4}/t^{4}}{k^{2}(1-\eta)}\,
\end{equation}
and the asymmetry parameters $\calE_{a}$ are related to asymmetry angles $\theta_{a}$ as
\begin{equation}\label{eq:epsilon_theta}
  e^{2\pi\calE_a}=\frac{\sin(\theta_a+\pi\Delta_a)}{\sin(\theta_a-\pi\Delta_a)}\zeta,\quad e^{-2i\theta_a}=\frac{\sin\pi(i\tilde{\calE}_a+\Delta_a)}{\sin\pi(i\tilde{\calE}_a-\Delta_a)},
\end{equation}
 where $\tilde{\calE}_{b}=\calE_{b},\tilde{\calE}_{f}=\calE_{f}+1/2$.

Thus all parameters of the IR solution are fully determined by the two asymmetry angles $\theta_f$ and $\theta_b$. We can relate these angles to the boson and fermion densities by Luttinger constraints.
According to \cite{GPS01,Gu:2019jub}, for a SYK-type model with UV source $\sigma$ we can define a U(1) charge at zero temperature
\begin{equation}\label{}
  \mathcal{Q}=-\int_{-\infty}^{\infty}\rd \tau \tau\sigma(\tau)G(-\tau) \,.
\end{equation}
For the case of our model, $\sigma(\tau)=\delta'(\tau)-\mu \delta(\tau)$, which yields
\begin{equation}\label{}
  \mathcal{Q}=\frac{G(0^+)+G(0^-)}{2}.
\end{equation}
Matching this with the definition of Green's functions, we get
\begin{equation}
   \mathcal{Q}_f = \left\langle f^{\dagger}f\right\rangle-\frac{1}{2}, \quad \mathcal{Q}_b = -\left\langle b^\dagger b\right\rangle-\frac{1}{2}\,.
\end{equation}
Also the charge can be calculated entirely from the IR asymptotics, related to $\theta$-parameter by
\begin{equation}\label{}
  \mathcal{Q}(\theta)=-\frac{\theta}{\pi}-\left(\frac{1}{2}-\Delta\right)\frac{\sin(2\theta)}{\sin(2\pi \Delta)}\,.
\end{equation}
Therefore, from (\ref{lutt1}), $\theta_f,\theta_b$ satisfy the following Luttinger constraints \cite{GPS01,Gu:2019jub}
\begin{eqnarray}
  \frac{\theta_f}{\pi}+\left(\frac{1}{2}-\Delta_f\right)\frac{\sin(2\theta_f)}{\sin(2\pi\Delta_f)} &=& \frac{1}{2}-\kappa+kp, \label{eq:Ltf}\\
  \frac{\theta_b}{\pi}+\left(\frac{1}{2}-\Delta_b\right)\frac{\sin(2\theta_b)}{\sin(2\pi\Delta_b)} &=& \frac{1}{2}+p\,. \label{eq:LtB}
\end{eqnarray}

\subsection{Bosonic Spinon + Fermionic holon}
\label{sec:bosonicspinons}

   We also discuss the other fractionalization scheme, consisting of bosonic spinons and fermionic holons,
\begin{equation}
  c_{il\alpha} = \fb_{i\alpha}\ff_{il}^\dagger, \quad
  S_{i\alpha\beta} = \fb_{i\alpha}^\dagger \fb_{i\beta}\,,
\end{equation}
 with the gauge charge constraint
 \begin{equation}\label{}
   \sum_{\alpha}\fb^{\dagger}_{i\alpha}\fb_{i\alpha}+\sum_{l}\ff_{il}^{\dagger}\ff_{il}=\kappa M,
 \end{equation} and filling density
 \begin{equation}\label{}
   \frac{1}{M'}\sum_{l}\langle\ff_{il}^{\dagger}\ff_{il}\rangle=p\,.
 \end{equation}
But most of our analysis will be carried out in the fractionalization scheme in Section~\ref{sec:fermionicspinons}.
Following the similar discussion of the other scheme, we can now derive the action to be
\begin{alignat}{1}\label{}
  S[\ff,\fb,\lambda] & = \int_0^{1/T}\rd \tau \left[\sum_{\alpha}\fb^{\dagger}_{\alpha}\left(\frac{\partial}{\partial \tau}+i\lambda\right)\fb_{\alpha}+\sum_{\alpha}\ff_{l}\left(\frac{\partial}{\partial \tau}+\epsilon_0+i\lambda\right)\ff_{l}-i\lambda\frac{M}{2}\right] \nonumber \\
   & +\frac{t^2}{M}\sum_{l\alpha}\int_{0}^{1/T}\rd \tau \rd \tau' \ff(\tau)\fb_\alpha^\dagger(\tau)G_c(\tau,\tau')\fb_{\alpha}(\tau')\ff^\dagger_{l}(\tau')\nonumber \\
   & -\frac{J^2}{2M}\sum_{\alpha\beta}\int_{0}^{1/T}\rd \tau \rd \tau' \fb_{\alpha}^\dagger(\tau)\fb_{\beta}(\tau)Q(\tau,\tau')\fb^\dagger_{\beta}(\tau')\fb_{\alpha}(\tau')\,.
\end{alignat}
The self consistency relations are
\begin{alignat}{2}\label{}
  G_{e}(\tau,\tau') & =\frac{-1}{MM'}\sum_{l\alpha}\langle c_{l\alpha}(\tau)c^\dagger_{l\alpha}(\tau')\rangle &&=\frac{-1}{M M'}\sum_{l\alpha}\left\langle \fb_{\alpha}(\tau)\ff^\dagger_l(\tau)\ff_l(\tau')\fb^\dagger_\alpha(\tau')\right\rangle\,, \\
  Q(\tau,\tau') &=\frac{1}{M^2}\sum_{\alpha\beta}\left\langle S_{\beta\alpha}(\tau)S_{\alpha\beta}(\tau')\right\rangle &&=\frac{1}{M^2}\sum_{\alpha\beta}\left\langle\fb^\dagger_\beta(\tau)\fb_\alpha(\tau)\fb^\dagger_\alpha(\tau')\fb_\beta(\tau')\right\rangle\,.
\end{alignat}
Introducing bilocal fields
\begin{equation}
  G_\fb(\tau,\tau') = \frac{-1}{M}\sum_{\alpha}\fb_{\alpha}(\tau)\fb^\dagger_{\alpha}(\tau'), \quad
  G_\ff(\tau,\tau') = \frac{-1}{M'}\sum_{l}\ff_l(\tau)\ff^\dagger_l(\tau')\,,
\end{equation}
and the corresponding self-energies $\Sigma_\fb(\tau,\tau')$, $\Sigma_\ff(\tau,\tau')$, we can obtain the saddle point equations
\begin{eqnarray}
  G_\fb(i\omega_n) &=& \frac{1}{i\omega_n+\mu_\fb-\Sigma_\fb(i\omega_n)}\,, \\
  G_\ff(i\omega_n) &=& \frac{1}{i\omega_n+\mu_\ff-\Sigma_\ff(i\omega_n)}\,, \\
  \Sigma_\fb(\tau) &=& -k t^2 G_\ff(\tau)G_\ff(-\tau)G_\fb(\tau)+J^2 G_\fb(\tau)^2 G_\fb(-\tau)\,, \\
  \Sigma_\ff(\tau) &=& t^2G_\ff(\tau)G_\fb(-\tau)G_\fb(\tau)\,,
\end{eqnarray} where we have substituted the self consistent relations
\begin{equation}
  G_{e}(\tau) = G_\ff(-\tau)G_\fb(\tau), \quad
  Q(\tau) = G_\fb(\tau)G_\fb(-\tau)\,.
\end{equation}
The chemical potentials should be adjusted to satisfy
\begin{equation}\label{}
  \left\langle \ff^{\dagger}\ff\right\rangle=p, \quad \left\langle\fb^\dagger \fb\right\rangle=\kappa-kp\,.
\end{equation}
Again we assume a conformal ansatz for the solution
\begin{equation}
 \quad G^{c}_{a}(\tau) =
-\begin{pmatrix}
e^{\pi \calE_{a} } \\
\zeta e^{-\pi \calE_{a}}
\end{pmatrix} \frac{b_{a}^{\Delta_{a}}}{|J\tau|^{2\Delta_{a}}},\quad G_{a}^{c}(i\omega) = -\frac{i C_{a}}{J}\begin{pmatrix}
e^{-i\theta_{a}}\\
- e^{i\theta_{a}}
\end{pmatrix} |\omega/J|^{2\Delta_{a}-1}\,,
\end{equation}
where $c$ in the superscript means conformal and subscript $a=\fb,\ff$ indexes boson and fermion respectively.
Assuming $\Delta_\ff=\Delta_\fb=1/4$, the coefficients $C_\fb$ and $C_\ff$ satisfy
\begin{equation}
  \begin{split}
  &(t^2/J^{2}) C_\ff^2 C_\fb^2 \cos(2\theta_\fb) = -\pi\,, \\
  &C_\fb^4 \cos(2\theta_\fb) -k(t^2/J^2) C_\fb^2 C_\ff^2 \cos(2\theta_\ff)= -\pi
 \end{split}
\end{equation}
and we find
\begin{equation}
C_{\fb} = \Big(\frac{\pi (1-\eta)}{-\cos(2\theta_{\fb})}\Big)^{1/4}, \quad C_{\ff} = \frac{J}{t}\Big(\frac{\pi }{-\cos(2\theta_{\fb})(1-\eta)}\Big)^{1/4}, \quad \eta \equiv - k\frac{\cos(2\theta_{\ff})}{\cos(2\theta_{\fb})}\,.
\end{equation}
 The unitarity constraints are $-\pi\Delta_\ff<\theta_\ff<\pi\Delta_\ff$, $\pi\Delta_\fb<\theta_\fb<\pi/2$ and
the condition for real $C_{\fb}$ and $C_{\ff}$ is $\eta <1$ as before.

The Luttinger constraints for $\theta_\ff,\theta_\fb$ are
\begin{eqnarray}
  \frac{\theta_\ff}{\pi}+\left(\frac{1}{2}-\Delta_\ff\right)\frac{\sin(2 \theta_\ff)}{\sin(2\pi\Delta_\ff)} &=& \frac{1}{2}-p\,, \label{eq:Ltff} \\
  \frac{\theta_\fb}{\pi}+\left(\frac{1}{2}-\Delta_\fb\right)\frac{\sin(2 \theta_\fb)}{\sin(2\pi\Delta_\fb)} &=& \frac{1}{2}+\kappa-kp\,. \label{eq:Ltfb}
\end{eqnarray}

If we ignore the Luttinger constraints, the saddle point equations above are identical to that of the other fractionalization scheme by substitution
\begin{equation}\label{eq:match}
  G_f(\tau)\leftrightarrow G_\fb(\tau),\quad G_b(\tau)\leftrightarrow G_\ff(\tau),\quad 0<\tau<1/T\,,
\end{equation}
and we transform $\tau<0$ into $0<\tau<\beta$ using KMS relation. In particular, when $\Delta_\ff=\Delta_b=\Delta_\fb=\Delta_f=1/4$, we could identify the Green's functions in the two schemes by
\begin{equation}\label{eq:thetarelation}
  \theta_{\ff}=\frac{\pi}{2}-\theta_b,\quad \theta_\fb=\frac{\pi}{2}-\theta_f,\quad C_\ff=C_b,\quad C_\fb=C_f.
\end{equation}
However, we point out that \eqref{eq:thetarelation} is incompatible with Luttinger constraints. For example, we can add \eqref{eq:LtB} and \eqref{eq:Ltff} together and using \eqref{eq:thetarelation} to see $\sin(2\theta_b)=1$, which leads to $p=0$, but \eqref{eq:Ltfb} has no solution for $p=0$. This reflects fact that at large $M$ limit the two fractionalization schemes are two different theories, and they are identical only at $M=2,M'=1$. 

\section{Operator Spectrum and Corrections to Conformality}\label{sec:OperSpectrum}

\subsection{Corrections to Conformality}
    The saddle point solutions $G_{f}$ and $G_{b}$ of the $t$-$J$ model can be viewed as a correlation functions in conformal field theory, which is deformed by infinite series of  irrelevant operators \cite{Gross:2016kjj,Klebanov:2016xxf,Klebanov:2018fzb}. Therefore the two-point functions receive corrections from these operators. We shall focus here on the fermionic spinon $+$ bosonic holon model and the results can be easily generalized to the other case. The general expression for the two-point function at zero temperature  reads
\begin{align}
\label{Gallcorr}
G(\tau) &=G^{c}(\tau)  \left(1-\sum_{h}^{}\frac{\alpha_{h} v_{h}}{|J\tau|^{h-1}}-\sum_{h,h'}^{}\frac{a_{hh'}\alpha_{h}\alpha_{h'}}{|J\tau|^{h+h'-2}}-\sum_{h,h',h''}^{}\frac{a_{hh'h''}\alpha_{h}\alpha_{h'}\alpha_{h''}}{|J\tau|^{h+h'+h''-3}}-\dots\right)\,,
\end{align}
where $v_{h}$, $a_{hh'}$, $a_{hh'h''}$, etc are  four-component vectors in $b/f$ and $+/-$ basis, so $v_{h}= (v_{hb+}, v_{hb-}, v_{hf+}, v_{hf-})^{\textrm{T}}$. For instance the fermionic Green's function for $\tau > 0$ reads
\begin{equation}
    G_{f}(\tau) = G^{c}_{f}(\tau)\left(1-\sum_{h}^{}\frac{\alpha_{h} v_{hf+}}{|J\tau|^{h-1}}-\sum_{h,h'}^{}\frac{a_{hh'f+}\alpha_{h}\alpha_{h'}}{|J\tau|^{h+h'-2}}-\sum_{h,h',h''}^{}\frac{a_{hh'h''f+}\alpha_{h}\alpha_{h'}\alpha_{h''}}{|J\tau|^{h+h'+h''-3}}-\dots\right)\,.
\end{equation}
The real numbers $\alpha_{h}$  are the same for all components $b/f$ and $+/-$. We are able to find these numbers only numerically. On the other hand the  vectors $v_{h}$ and  anomalous dimensions $h$ can be found analytically from solving the equation
\begin{equation}
K_{G}(h)v_{h} = v_{h}\,, \label{eq:KGunit}
\end{equation}
where the $4\times 4$ matix $K_{G}(h) = W_{\Sigma}(h) W_{G}(h)$ is a product of two matrices $W_{\Sigma}(h)$ and $W_{G}$ which in the $(b/f) \times (+/-)$ basis are given by formulas
\begin{equation}\label{eq:Wsigmaexpression}
    W_\Sigma(h)=\frac{\sec(\pi h)}{(1-2h)}
    \begin{pmatrix}
    \sin(\pi h+2\theta_{b}) & \sin(2\theta_{b})-1 \\
    -\sin(2\theta_{b})-1 & \sin(\pi h-2\theta_{b})
    \end{pmatrix} \oplus \begin{pmatrix}
    \sin(\pi h+2\theta_{f}) & \sin(2\theta_{f})-1 \\
    -\sin(2\theta_{f})-1 & \sin(\pi h-2\theta_{f})
    \end{pmatrix},
\end{equation}
\begin{equation}\label{}
 W_G=\left(
\begin{array}{cccc}
 1 & 0 & 1 & 1 \\
 0 & 1 & 1 & 1 \\
 \eta  & \eta  & 2-\eta & 1-\eta \\
 \eta  & \eta  & 1-\eta & 2-\eta \\
\end{array}
\right),
\end{equation}
and we remind that $\eta=- k \cos(2\theta_b)/\cos(2\theta_f)$.  We notice that the matrix $W_{\Sigma}(h)$ is block-diagonal in $b/f$ space and $W_{G}$ does not depend on $h$. For $\eta =0$ the fermionic sector is independent on the bosonic one and its Green's function is exactly the same as in the complex SYK model.

The results described above are based on the Kitaev-Suh theory \cite{kitaev2017, Guo:2020aog, Tikhanovskaya:2020elb}.
     The derivation of the kernel $K_{G}$ is done in \cite{Guo:2020aog}.
     The kernel $K_{G}$ for bosonic spinon + fermionic holon case is obtained by substitution $\theta_f\to \theta_\fb$, $\theta_b\to \theta_\ff$, for details see Appendix \ref{app:kernel}. The coefficients $a_{hh'}$, $a_{hh'h''}$ can be calculated using the recursion procedure described in \cite{Tikhanovskaya:2020elb}. We present explicit derivation of $a_{hh'}$ for the $t$-$J$ model in Appendix \ref{app:kernel}.

 The linear order zero-temperature correction \eqref{Gallcorr} was generalized to finite temperature in \cite{Guo:2020aog, Tikhanovskaya:2020elb}. The result takes the form on the interval $\tau \in (0,\beta)$:
 \begin{equation}\label{GcorrfinitT}
     G_a(\tau)=G^c_a(\tau)\left(1-\frac{v_{ha+}+v_{ha-}}{2}\frac{\alpha_h}{(\beta J)^{h-1}}f_h^\mathrm{A}(\tau)-\frac{v_{ha+}-v_{ha-}}{2}\frac{\alpha_h}{(\beta J)^{h-1}}f_h^\mathrm{S}(\tau)\right)\,,
 \end{equation} where $G^c_a(\tau)$ is the finite temperature conformal Green's function
\begin{equation}
    G^{c}_a(\tau) = -b_a^{\frac{1}{4}} \left(\frac{\beta J}{\pi} \sin \frac{\pi \tau}{\beta} \right)^{-\frac{1}{2}} e^{2\pi \calE_a (\frac{1}{2}-\frac{\tau}{\beta})}\,,
\end{equation}
and the functions $f_{h}^{\textrm{A}/\textrm{S}}(\tau)$ are given by the formula
 \begin{align}
f_{h}^{\textrm{A}}(\tau) &= \frac{(2\pi)^{h-1} \Gamma(h)^{2}}{\Gamma(2h-1)\cos(\frac{\pi h}{2})}  \left(A_{h}(e^{i\frac{2\pi \tau}{\beta}})+A_{h}(e^{-i\frac{2\pi \tau}{\beta}})\right)\,, \notag\\
f_{h}^{\textrm{S}}(\tau) &= \frac{(2\pi)^{h-1} \Gamma(h)^{2}}{\Gamma(2h-1)\sin(\frac{\pi h}{2})}  \left(iA_{h}(e^{i\frac{2\pi \tau}{\beta}})-iA_{h}(e^{-i\frac{2\pi \tau}{\beta}})\right)\,,
\label{eq:fhAfhS}
\end{align}
 where  $A_{h}(u)=(1-u)^{h}\mathbf{F}(h,h,1;u)$ and
 $\mathbf{F}$ is the regularized hypergeometric function. We stress that there are no vectors in the formula (\ref{GcorrfinitT}) and it is correct for both bosons and fermions $a=b,f$ with their corresponding conformal two-point functions $G^{c}(\tau)$ and components $v_{h\pm}$.

\subsection{Operator Spectrum}

In this subsection we describe solutions $h$ of the equation (\ref{eq:KGunit}).   Such values of $h$ correspond to scaling dimensions of the bilinear operators $\mathcal{O}_{h}$ in the $t$-$J$ model \cite{Tikhanovskaya:2020elb}.  We can rewrite the equation for $h$ in the form
\begin{equation}\label{eq:det(1-K)=0}
  \det(1-K_G(h))=0\,.
\end{equation}
We  label positive solutions of the above equation by $h_{0}, h_{1}, h_{2}, \dots $ and notice that the sums in (\ref{Gallcorr}) are taken over this series. Several remarks are in order. First we notice that there are always two $h=1$ solutions which are related to $U(1)$ global symmetry and $U(1)$ gauge symmetry. The $h=1$ modes are not included in the series $h_{0}, h_{1}, h_{2},\dots$. The effect of these modes is already taken into account in the asymmetry parameters $\calE_{f}$ and $\calE_{b}$ in $G_{f}^{c}$ and $G_{b}^{c}$ \cite{Gu:2019jub, Guo:2020aog}.  Second there is another parameter-invariant solution $h=2$ which corresponds to emergent time-reparametrization symmetry of the Eqs.\eqref{Eq:EoM1}-\eqref{Eq:EoM4}. We label this solution by $h_{0}$, so we always have $h_{0}=2$. The rest of the modes are ordered and their values  depend on $k,\theta_f,\theta_b$, so we have
\begin{equation}
h_{0}=2, \quad h_{1} \leq h_{2} \leq h_{3} \leq h_{4} \leq  \dots\,.
\end{equation}
We  will  use notation $\alpha_{i}$ for $\alpha_{h_{i}}$ and $v_{i}$ for $v_{h_{i}}$. The eigenvector $v_{0}$ has a simple form
 \begin{equation}\label{eq:v0A}
   v_{0}=\begin{pmatrix}
             1-\frac{3}{2}\sin(2\theta_b) \\
             1+\frac{3}{2}\sin(2\theta_b) \\
             1-\frac{3}{2}\sin(2\theta_f) \\
             1+\frac{3}{2}\sin(2\theta_f)
           \end{pmatrix}
 \end{equation}
 and  is not normalized to unity. We assume that all other vectors are normalized to unity, so $v_{i}^{\textrm{T}}v_{i}=1$ for $i=1,2,3,\dots$.

 Our interest is in modes which can be potentially smaller than the time-reparameterization mode  $h_{0}=2$. In the $t$-$J$ model the mode $h_{1}$ is less than $3/2$ for arbitrary parameters and  thus it is always more relevant than $h_{0}=2$ mode. In addition, the mode $h_{2}$ can become smaller than $h_{0}$ for some regions of parameters.
  A qualitative plot of these potentially relevant operators for the fermionic spinon case of Section~\ref{sec:fermionicspinons} is shown in Fig.~\ref{fig:diagram_fermionic_spin} (the derivation is in Appendix.~\ref{app:diagram}).
 \begin{figure}
  \centering
  \includegraphics[width=15cm]{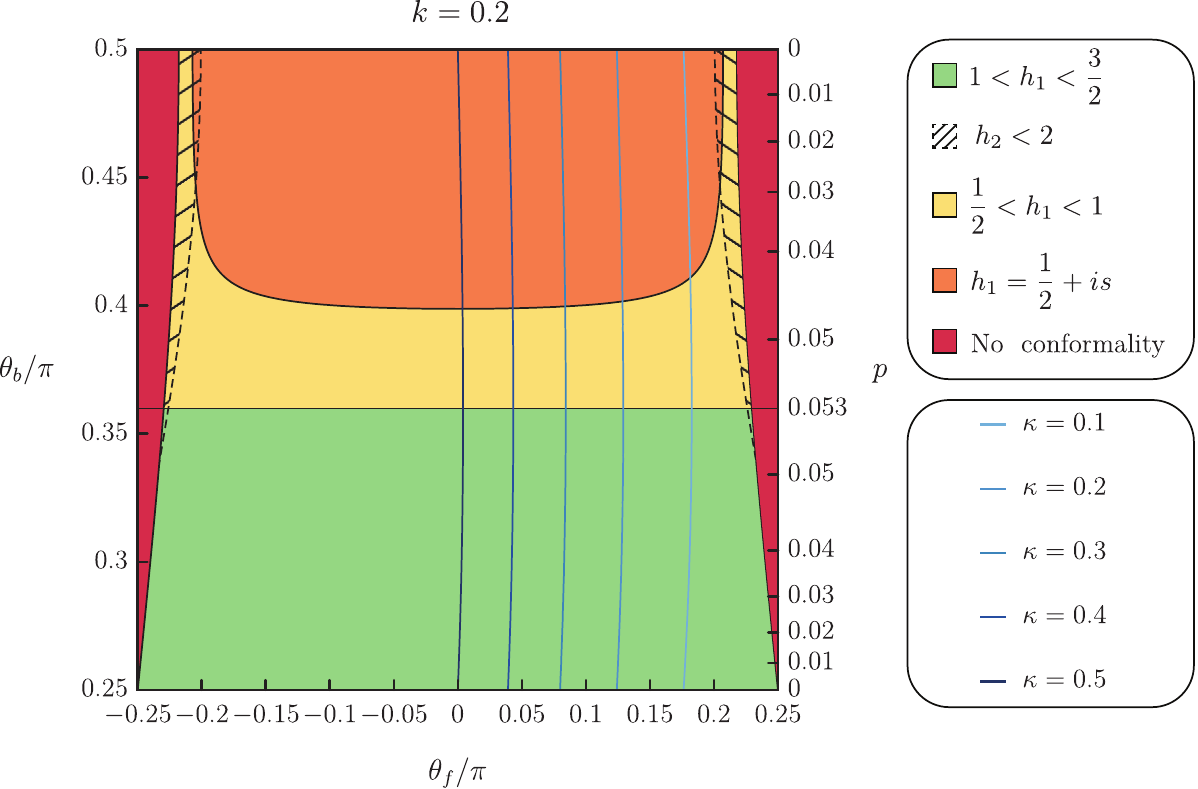}
  \caption{Scaling dimensions $h$ of  potentially dominant operators in the $(\theta_f,\theta_b)$ plane. Here $k=0.2$. Regions corresponding to different ranges of $h$ are coded with different colors, with boundaries given by Eqs.\eqref{eq:existence1},\eqref{eq:bound1},\eqref{eq:bound2},\eqref{eq:bound3}. There is a region dubbed ``No conformality" meaning \eqref{eq:conformal_condition} is violated. On the right side, we also label different values of $p$ obtained from the Luttinger constraint \eqref{eq:LtB}. In particular, the maximal $p_{\textrm{max}}\approx 0.053$ is reached when $\theta_b=\frac{1}{2}\arccos\left(\frac{-2}{\pi}\right)$. The colored lines are constant $\kappa$ contours calculated from the Luttinger constraints \eqref{eq:Ltf},\eqref{eq:LtB}. }\label{fig:diagram_fermionic_spin}
\end{figure}
We remind the readers that $\theta_f$ and $\theta_b$ are constrained by unitarity $-\pi/4<\theta_f<\pi/4$, $\pi/4<\theta_b<\pi/2$. The mode $h_1$ (see Fig.~\ref{fig:hsplot})  has $h_1=3/2$ at zero doping $p=0$, and its scaling dimensions decreases as $p$ grows, and it can become relevant $h_1<1$ or complex $h_1=1/2+is,~s\in \mathbf{R}$.
  \begin{figure}[h!]
  \centering
  \includegraphics[width=10cm]{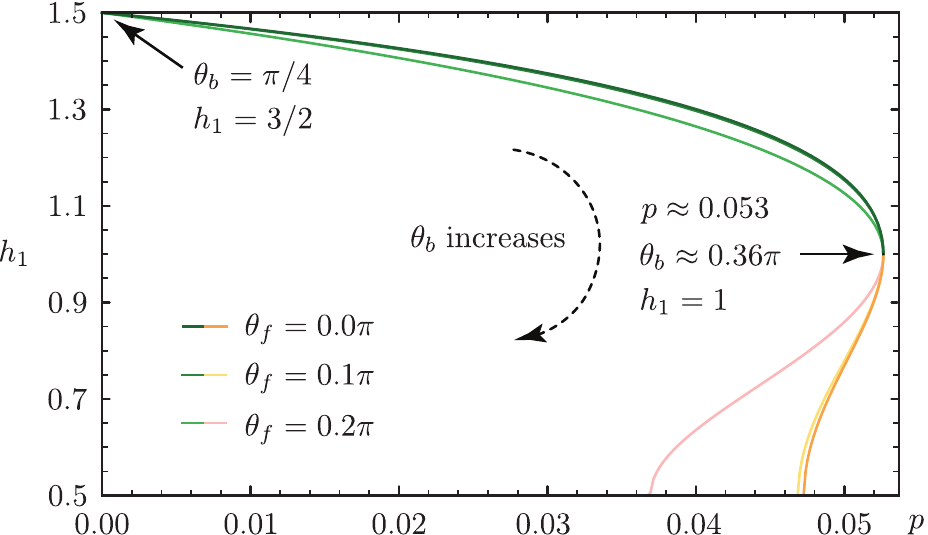}
  \caption{The scaling dimension $h_1$   plotted versus doping density $p$. The plot is obtained by taking three vertical cuts from Fig.~\ref{fig:diagram_fermionic_spin} at $\theta_f=0,0.1\pi,0.2\pi$. The scaling dimension $h_1=3/2$ when $p=0$ and reaches $h_1=1$ when $p=p_{\textrm{max}}\approx 0.053$. Then $h_1$ starts to decrease with $p$ until it reaches $h_1=1/2$ and becomes complex. Here we used $k=0.2$.}\label{fig:hsplot}
\end{figure}
 At first glance, it seems that the low-energy properties would be dominated by this operator. However, as we will see in Section~\ref{sec:Numerics}, the prefactor $\alpha_{1}$ accompanying this mode is small compared to $\alpha_{0}$ of the $h_{0}=2$ mode when doping $p$ is small. A partial explanation is the  coefficient $\alpha_{1}$ is proportional to $k_G'(h_1)^{-1}$ (see \eqref{eq:deltaGalpha}), and when $p$ is small, $h_1$ is close to the pole at $k_G(3/2)$, so $k_G'(h_1)$ is large. Therefore, we expect time reparameterization mode $h_{0}=2$ to be important above some threshold temperature $T>T_1(p)$, where $T_1(p)\to 0$ as $p\to 0$.
There is another potentially relevant operator $h_{2}$ which can move below 2, but it only appears in a restricted range of parameters.

\begin{figure}
  \centering
  \includegraphics[width=15cm]{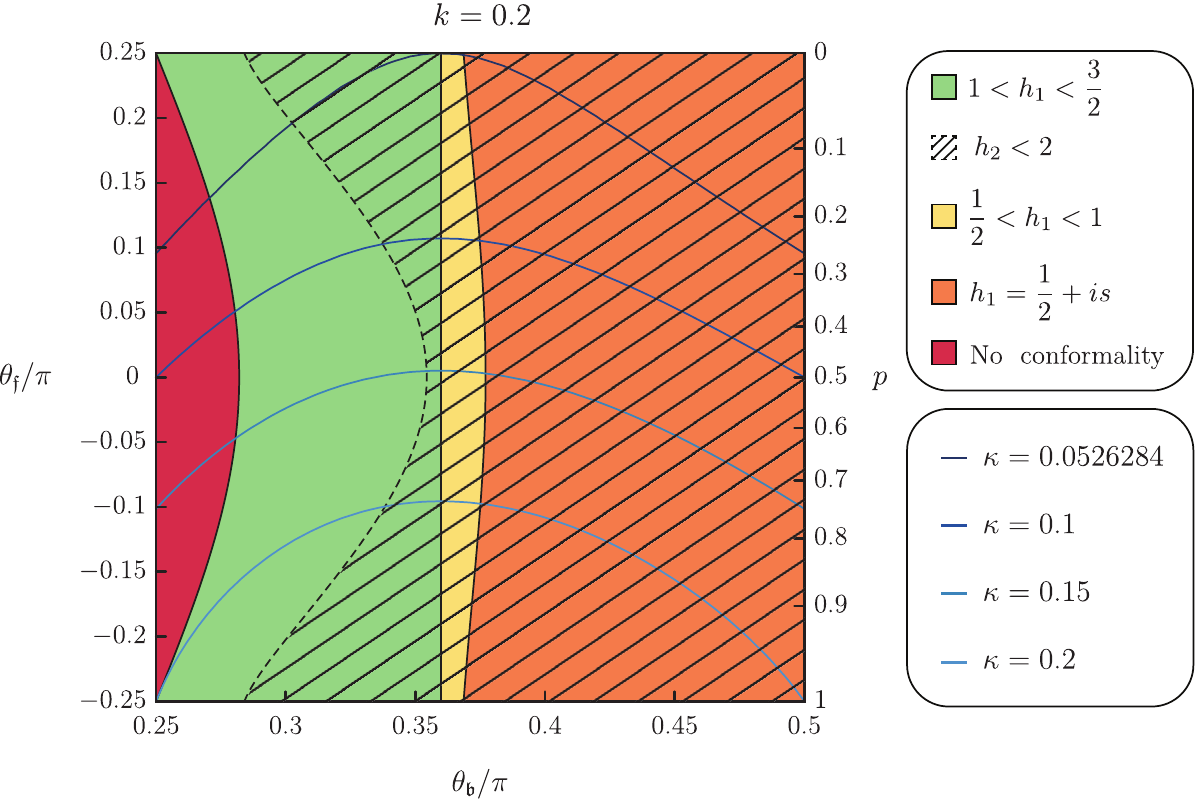}
  \caption{As in Fig.~\ref{fig:diagram_fermionic_spin}, but now for the bosonic spinon model. }\label{fig:bosonic_spin_plot}
\end{figure}
The analysis of conformal corrections and operator spectrum in the bosonic spinon $+$ fermionic holon case of Section~\ref{sec:bosonicspinons} is very similar to the fermionic spinon + bosonic holon model. The kernel $K_G$ can be obtained from the previous model by changing $\theta_f\to\theta_\fb$, $\theta_b\to\theta_\ff$ (see Appendix.~\ref{app:kernel}).
The discussions of operators at $h=1$ and $h=2$ in the previous subsection still applie here. The analysis of other potentially relevant operators is shown in Fig.~\ref{fig:bosonic_spin_plot}.
In this case the second potentially relevant operator $h_{2}$ appears in a larger region of parameter space.

\section{Gauge Invariant Observables}\label{sec:GIO}

In this section we present results for some gauge invariant observables including electron spectral density $\rho_{e}(\omega)$, spin spectral density $\rho_Q(\omega)$ and optical conductivity $\Re ~\sigma(\omega)$. We shall be focusing here on the fermionic spinon + bosonic holon fractionalization of Section~\ref{sec:fermionicspinons}.

\subsection{Electron Spectral Density}

  The electron Green's function is a product of the spinon and holon Green's functions
\begin{equation}\label{GeasGfGb}
  G_{e}(\tau)=-G_f(\tau)G_b(-\tau)\,.
\end{equation}
  To obtain the electron spectral density, we shall first analytically continue to retarded-Green's function from imaginary time to the real one
\begin{equation}\label{eq:GeRfromGtau}
  G_{e,R}(t)=i\theta(t) (G_{e}(it+0)-G_{e}(it-0))\,,
\end{equation}
where $\theta(t)$ is the Heaviside step function.
Then we can find $G_{e,R}(\omega)$ in the frequency space and extract the spectral density as
\begin{equation}\label{eq:rhoefromGeR}
  \rho_{e}(\omega)=-\frac{1}{\pi}\Im G_{e,R}(\omega)\,.
\end{equation}
The electron spectral density can also be written in terms of the spectral densities of spinon $\rho_f=-\frac{1}{\pi}\Im G_{f,R}$ and holon $\rho_b=-\frac{1}{\pi}\Im G_{b,R}$ as
\begin{equation}\label{}
  \rho_{e}(\Omega)=\int_{-\infty}^{+\infty} \rd \nu (n_F(\Omega+\nu)+n_B(\nu))\rho_f(\Omega+\nu)\rho_b(\nu)\,,
\end{equation}
where $n_{F/B}(\omega)=1/(e^{\beta \omega}-\zeta)$ is the Fermi or Bose distribution.
We will perform the calculation at both zero temperature at finite temperature. At zero temperature we include nonlinear corrections, and at finite temperature we will restrict expressions to linear order corrections.

\subsubsection{Zero temperature}

Using \eqref{Gallcorr} and (\ref{GeasGfGb}), we can write the electron Green's function as
\begin{equation}\label{}
  G_{e}(\tau)=G_{e}^c(\tau)\bigg(1-\sum_h \frac{\alpha_h}{|J\tau|^{h-1}}(v_{hf}+\bar{v}_{hb})-\sum_{hh'}\frac{\alpha_h\alpha_{h'}}{|J\tau|^{h+h'-2}}(a_{hh'f}+\bar{a}_{hh'b}-v_{hf}\bar{v}_{h'b})-\dots\bigg).
\end{equation} Here $v_{hf(b)}$ means taking the fermionic (bosonic) component of the 4-component vector $v_h$, and similar meanings apply to $a_{hh'f(b)}$. $\bar{v}_{hb}$ means switching the $\pm$ components of $v_{hb}$ and similar for $\bar{a}_{hh'b}$. $G_{e}^c(\tau)$ is the conformal electron Green's function
\begin{equation}\label{eq:get=0}
 \quad G^{c}_{e}(\tau) =
-\begin{pmatrix}
e^{\pi \calE_{e} } \\
-e^{-\pi \calE_{e}}
\end{pmatrix} \frac{C_{e}}{|J\tau|}\,, \quad C_{e} = \frac{J}{t}\Big(\frac{-\cos(2\theta_{b})}{4\pi}\Big)^{1/2}\,,
\end{equation}
where $\calE_{e} = \calE_{f}-\calE_{b}$ and we have used that $\Delta_{f}=\Delta_{b}=1/4$.
Analytically continuing to real-time and extract the spectral density, we obtain
\begin{equation}\label{eq:rhocT=0}
  \rho_{e}(\omega)=\rho_{e}^c(\omega)\bigg(1-\sum_h \frac{\alpha_h (v_{hf}+\bar{v}_{hb})}{\Gamma(h)}\left|\frac{\omega}{J}\right|^{h-1}
 -\sum_{hh'}\frac{\alpha_h\alpha_{h'} (a_{hh'f}+\bar{a}_{hh'b}-v_{hf}\cdot\bar{v}_{h'b})}{\Gamma(h+h'-1)}\left|\frac{\omega}{J}\right|^{h+h'-2}
  -\dots\bigg).
\end{equation} Here $\rho_{e}^c(\omega)$ is the electron-spectral weight of the conformal solution
\begin{equation}\label{eq:rhocConf}
  \rho_{e}^c(\omega)=\frac{C_{e}}{J}\begin{pmatrix}
e^{\pi \calE_{e} } \\
e^{-\pi \calE_{e}}
\end{pmatrix}\,,
\end{equation}
where $C_{e}$ is defined in (\ref{eq:get=0}).

\subsubsection{Finite Temperature}

  We will work on the interval $0<\tau<\beta$. The electron Green's function to linear order in $\alpha_{h}$ is
   \begin{align}
\label{GecorrfinitT}
G_{e}(\tau) =& G_{e}^{c}(\tau) \bigg(1-\sum_{h}^{}\frac{\alpha_{h}}{2(\beta J)^{h-1}}\big((v_{hf+}+v_{hf-}+v_{hb+}+v_{hb-})f_{h}^{\textrm{A}}(\tau)\notag\\
&+(v_{hf+}-v_{hf-}-v_{hb+}+v_{hb-})f_{h}^{\textrm{S}}(\tau)\big)-\dots\bigg)\,,
\end{align}
  where the conformal electron Green's function at finite temperature is
\begin{equation}\label{eq:Gcc}
  G_{e}^c(\tau)=-C_{e}\Big(\frac{\beta J}{\pi}\sin\frac{\pi\tau}{\beta}\Big)^{-1}e^{2\pi\calE_{e}(\frac{1}{2}- \frac{\tau}{\beta})}\,,
\end{equation}
and the functions $f_{h}^{\textrm{A}}(\tau)$ and $f_{h}^{\textrm{S}}(\tau)$ are given in (\ref{eq:fhAfhS}).

To find electron spectral density at finite temperature we analytically continue $G_{e}(\tau)$ to real
time using (\ref{eq:GeRfromGtau}) and then take the Fourier transform to find $G_{e,R}(\omega)$ in the frequency space.
Finally using (\ref{eq:rhoefromGeR}) we find for the  electron spectral density
\begin{align}
\rho_{e}(\omega)=&\rho_{e}^{c}(\omega)\bigg(1- \sum_{h} \frac{\alpha_{h}}{2 (\beta J)^{h-1}} \Big((v_{hf+}+v_{hf-}+v_{hb+}+v_{hb-})\mathcal{R}_{h}^{\textrm{A}}\big( \frac{\beta \omega}{2\pi }-\calE_{e}\big)\notag\\
&+(v_{hf+}-v_{hf-}-v_{hb+}+v_{hb-})\mathcal{R}_{h}^{\textrm{S}}\big( \frac{\beta \omega}{2\pi }-\calE_{e}\big)\Big)-\dots\bigg)\,,
\label{eq:rhoefinal}
\end{align}
where the conformal electron spectral density is
\begin{equation}\label{eq:rhocc}
\rho_{e}^c(\omega)=\frac{C_{e}\cosh(\frac{\beta\omega}{2})}{J\cosh(\frac{\beta\omega}{2}-\pi\ce_{e})}\,,
\end{equation}
and the functions $\mathcal{R}_{h}^{\textrm{A}}(\omega)$ and $\mathcal{R}_{h}^{\textrm{S}}(\omega)$ are
\begin{equation} \label{eq:RAandRS}
\begin{split}
\mathcal{R}_{h}^{\textrm{A}}(\omega) &=  \frac{2 \left(\frac{\pi }{2}\right)^h \Gamma (h)}{\sqrt{\pi } \sin \left(\frac{\pi  h}{2}\right) \Gamma \left(h-\frac{1}{2}\right)}\textrm{Re}\,\pFq{3}{2}{h,1-h,\frac{1}{2}+i\omega}{1, 1}{1}\,, \\
\mathcal{R}_{h}^{\textrm{S}}(\omega) &=  \frac{2 \left(\frac{\pi }{2}\right)^h \Gamma (h)}{\sqrt{\pi } \cos \left(\frac{\pi  h}{2}\right) \Gamma \left(h-\frac{1}{2}\right)}\textrm{Im}\,\pFq{3}{2}{h,1-h,\frac{1}{2}+i\omega}{1, 1}{1}\,,
\end{split}
\end{equation}
where $\,_{3}\textbf{F}_{2}$ is the regularized hypergeometric function.
For derivation of similar to (\ref{eq:rhoefinal}) results see \cite{Tikhanovskaya:2020elb}.
Retaining only $h_0=2$ mode in (\ref{eq:rhoefinal})  we find
\begin{align}
&\rho_{e} (\omega) = \frac{C_{e} \cosh(\frac{\beta\omega}{2})}{J\cosh(\frac{\beta\omega}{2}-\pi\calE_{e})}
\left(1-\frac{\pi \alpha_{0}}{\beta J}\Big(\frac{\beta\omega}{2\pi}-\calE_{e}\Big)\Big(4\tanh(\frac{\beta\omega}{2}-\pi\calE_{e}) -3(\sin(2\theta_f)-\sin(2\theta_b))\Big)\right)\,,\label{eq:rhoeresult}
\end{align}
where we have used the explicit expression \eqref{eq:v0A} for the eigenvector $v_{0}$.

\subsection{Conductivity}

The Kubo formula for conductivity on Bethe lattice is derived in \cite{Guo:2020aog}:
\beq
\Re~\sigma(\omega) = \frac{ 2\pi M' e^2 t^2 a^{2-d}}{z} \int_{-\infty}^{+\infty} \rd \Omega  \rho_{e} (\Omega)\rho_{e}(\Omega+\omega)\frac{n_F(\Omega)-n_F(\Omega+\omega)}{\omega}\,, \label{sigmaomega}
\eeq
where $a$ is lattice constant and $d$ is spatial dimension. We can evaluate the expression in three scenarios as below:

\subsubsection{Zero temperature Optical conductivity}

  At zero temperature, we can evaluate the conductivity to nonlinear order in $\alpha_h$ using \eqref{eq:rhocT=0}, the result is
  \begin{equation}\label{eq:sigmaT=0}
  \begin{split}
    &\Re~\sigma(\omega)=\sigma_0\Bigg[1-\sum_h \frac{2 \alpha_h}{\Gamma(h+1)}(v_{hf+}+v_{hf-}+v_{hb+}+v_{hb-})|\omega/J|^{h-1}-\sum_{hh'}\frac{2\alpha_h\alpha_{h'}|\omega/J|^{h+h'-2}}{\Gamma(h+h')}\\
    &\times[a_{hh'f+}+a_{hh'f-}+a_{hh'b+}+a_{hh'b-}-v_{hf+}v_{h'b-}-v_{hf-}v_{h'b+}-(v_{hf+}+v_{hb-})(v_{h'f-}+v_{h'b+})/2]\Bigg].
  \end{split}
  \end{equation} Here $\sigma_0$ is the $T=0$ DC conductivity given by
\begin{equation}\label{eq:sigma0T0}
  \sigma_0=-\frac{M' e^2 a^{2-d} \cos(2\theta_{b})}{2z}\,,
\end{equation}
where we used (\ref{eq:rhocConf}) for $\rho_{e}^{c}(\Omega)$.

\subsubsection{Finite temperature DC conductivity}

    The finite temperature DC conductivity can be evaluated by
  \beq
\sigma_{DC} = \frac{ 2\pi M' e^2 t^2 a^{2-d}}{z} \int_{-\infty}^{+\infty} \rd \Omega  \rho_{e} (\Omega)^2\left(-\frac{\partial n_F}{\partial \Omega}\right)\,, \label{sigmaDC}
\eeq

\begin{figure}
  \centering
  \includegraphics[width=0.5\textwidth]{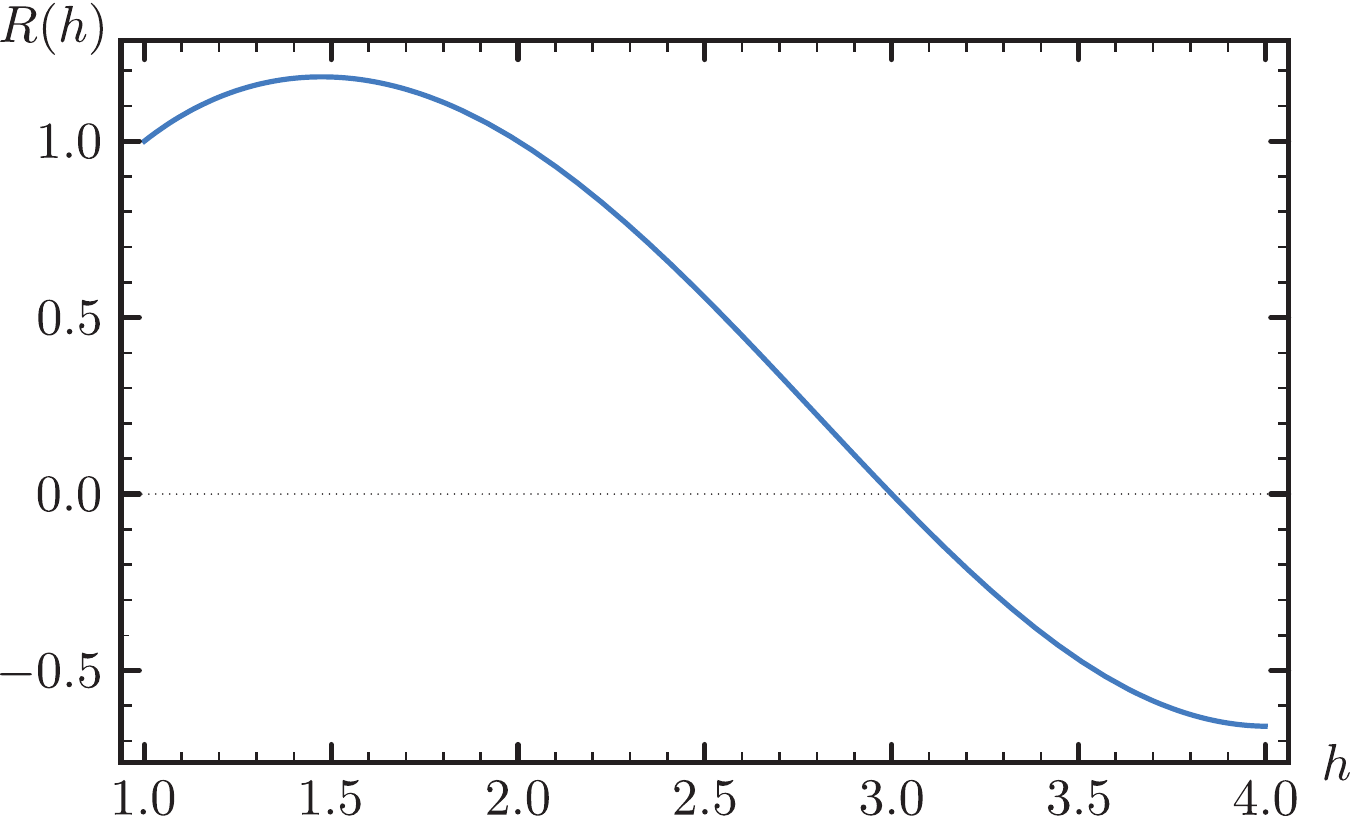}
  \caption{The function $R(h)$ defined in \eqref{eq:Rh}. Some special values are $R(1)=R(2)=1$.}\label{fig:Rplot}
\end{figure}

  The conformal contribution is given by
\begin{equation}\label{}
\begin{split}
  \sigma_{DC}^c&=\frac{ 2\pi M' e^2 t^2 a^{2-d}}{z}\int_{-\infty}^{+\infty} \rd \Omega \rho_{e}^c(\Omega)^2 \left(-\frac{\partial n_F}{\partial \Omega}\right) =\sigma_0\,,
\end{split}
\end{equation}
where for $\rho_{e}^{c}(\Omega)$ we used (\ref{eq:rhocc}).
We can also include conformal corrections up to linear order
\begin{equation}\label{}
  \delta_h \sigma_{DC}=\sigma_0 \frac{\int \rd \Omega\, 2 \rho_{e}^c(\Omega) \delta_h \rho_{e}(\Omega)\left(-\frac{\partial n_F}{\partial \Omega}\right)}{\int \rd \Omega \rho_{e}^c(\Omega)^2\left(-\frac{\partial n_F}{\partial \Omega}\right)}\,.
\end{equation}
Here $\delta_h \rho_{e}(\Omega)$ contains two terms given in \eqref{eq:rhoefinal}, so we have
\begin{equation}\label{}
  \frac{\delta_h \sigma_{DC}}{\sigma_0}=-\frac{\alpha_h}{(\beta J)^{h-1}}(v_{hf+}+v_{hf-}+v_{hb+}+v_{hb-}) R(h)\,,
\end{equation} where
\begin{equation}\label{eq:Rh}
  R(h)=\frac{\pi}{2}\int_{-\infty}^{+\infty} \rd \omega \frac{ \mathcal{R}_{h}^{\textrm{A}}(\omega)}{\cosh^2(\pi \omega)} =-\frac{2 \left(\frac{\pi }{2}\right)^{h-1} \cos \left(\frac{\pi  h}{2}\right) \Gamma (h-1)}{\sqrt{\pi } h \Gamma \left(h-\frac{1}{2}\right)}\,.
\end{equation}
It is clear from (\ref{eq:RAandRS}) that $\mathcal{R}_{h}^{\textrm{A/S}}(\omega)$ is an even/odd function of $\omega$, and therefore only  $\mathcal{R}_{h}^{\textrm{A}}$ contributes. Thus, after replacing variable $\omega = \frac{\beta \Omega}{2\pi}-\calE_{e}$ and taking the integral we arrive to (\ref{eq:Rh}).
A numerical plot of $R(h)$ is shown in Fig.~\ref{fig:Rplot}.

   Therefore the DC conductivity is
 \begin{equation}\label{eq:sigmaDCT}
   \sigma_{DC}=\sigma_0\left[1-\sum_{h}\alpha_h(v_{hf+}+v_{hf-}+v_{hb+}+v_{hb-}) R(h) \left(\frac{T}{J}\right)^{h-1} -\dots \right].
 \end{equation}
 In particular, from the $h_{0}=2$ mode we can recover the linear resistivity obtained in \cite{Guo:2020aog}
 \begin{equation}\label{}
   \rho_{DC}=\frac{1}{\sigma_0}\left(1+\frac{4\alpha_0 T}{J}-\dots\right),
 \end{equation} where we have used the eigenvector \eqref{eq:v0A}. There is a factor of two difference from \cite{Guo:2020aog} because we have normalized the eigenvector differently.

\subsubsection{Finite temperature optical conductivity at linear order}

  For the optical conductivity at finite temperature, we can analytically determine the contribution from the $\alpha_h$ correction. The result is
\begin{equation}\label{eq:sigmah=2}
  \Re \sigma(\omega)=\sigma_0\left(1-\sum_{h}\frac{\alpha_h}{(\beta J)^h}(v_{hf+}+v_{hf-}+v_{hb+}+v_{hb-})\Sigma_h\left(\frac{\beta \omega}{2\pi}\right)-\dots\right),
\end{equation} where
\begin{equation}\label{}
  \Sigma_h(x)=\frac{2\left(\frac{\pi}{2}\right)^{h}\Gamma(h)}{\sqrt{\pi}\sin\left(\frac{\pi h}{2}\right)\Gamma\left(h-\frac{1}{2}\right)}\Re \,\pFq{3}{2}{h,1-h,1-ix}{1, 2}{1}\,.
\end{equation}
In particular, for the $h_0=2$ mode we have
\begin{equation}\label{}
  \Re \sigma(\omega)=\sigma_0\left(1-\frac{2\alpha_0 \omega}{J}\coth(\frac{\beta\omega}{2})-\dots\right)\,.
\end{equation}
Here we have used the eigenvectors \eqref{eq:v0A}. The detail of the calculation is in Appendix~\ref{sec:sigmah=2}.

\subsection{Spin Spectral Density}

  In the large $M$ limit, the spin-spin correlator factorizes as $Q(\tau)=G_f(\tau)G_f(-\tau)$, and the spin spectral density is defined as $\rho_Q=-\frac{1}{\pi}\Im Q_{R}(\omega)$. It can be expressed in terms of spinon spectral weight $\rho_f$ as
  \begin{equation}
\rho_{Q}(\omega)=
 \int_{-\infty}^{\infty} d\nu \rho_f(\nu)\rho_f(\nu-\omega)(n_F(\nu-\omega)-n_F(\nu))\,.
\end{equation}
The computation of this quantity is the same as the previous paper \cite{Tikhanovskaya:2020elb}, and we report the results here.

  \subsubsection{Zero temperature}

  At zero temperature, the conformal spin spectral density is
\begin{equation}\label{eq:Qconf}
  \rho_Q^c(\omega)=\frac{C_Q}{J} \sgn \omega, \quad C_{Q} = \Big(\frac{\cos(2\theta_{f})}{4\pi}(1-\eta)\Big)^{1/2}\,.
\end{equation}
Including the conformal corrections, we have
\begin{align}\label{eq:rhoQT=0}
\rho_{Q}(\omega)= & \rho_{Q}^{c}(\omega)\bigg(1-\sum_{h} \frac{\alpha_{h}(v_{hf+}+v_{hf-})}{\Gamma(h)}\left|\frac{\omega}{J}\right|^{h-1}\notag\\
& -\sum_{h,h'}\frac{\alpha_h\alpha_{h'}(a_{hh'f+}+a_{hh'f-}-v_{hf+}v_{h'f-})}{\Gamma(h+h'-1)}\left|\frac{\omega}{J}\right|^{h+h'-2}-\dots\bigg)\,,
\end{align}

\subsubsection{Finite temperature}

 At finite temperature, we write the spin spectral weight to linear order in $\alpha_{h}$ as
\begin{equation}\label{}
  \rho_Q(\Omega)=\rho_Q^c(\omega)\bigg(1-\sum_{h}(v_{fh+}+v_{fh-}) \frac{\alpha_{h}}{(\beta J)^{h-1}} \mathcal{R}_{h}^{\textrm{A}}(\omega)-\dots\bigg)\,,
\end{equation}
where the conformal contribution to spin spectral weight is
\begin{align}
\rho_{Q}^{c}(\omega) &=\frac{C_{Q}}{J} \tanh (\frac{\beta \omega}{2})\,.
\end{align}
and the function $\mathcal{R}_{h}^{\textrm{A}}(\omega)$ defined in (\ref{eq:RAandRS}).
Considering only the correction from $h_0=2$ mode, we have
  \begin{align}\label{eq:rhoQT}
\rho_{Q}(\omega)= \frac{C_Q}{J} \tanh\big(\frac{\beta \omega}{2}\big)\left(1 -\frac{2\alpha_{0}\omega}{J} \tanh\big(\frac{\beta \omega}{2}\big)-\dots\right)\,.
\end{align}

\section{Numerical results for the $t$-$J$ model}
\label{sec:Numerics}

In this section we numerically study the saddle point equations at zero temperature for the $t-J$ model, and use the solutions to constrain the coefficients of the conformal perturbations. We find  that the contribution of the linear term from the $h_0 = 2$ mode dominates in the IR limit for the spin spectral density, but that the $h_1$ and $h_0$ modes are comparable in other observables.
We focus on the case of fermionic spinon and bosonic holon model discussed in the  Section~\ref{sec:fermionicspinons}. The solution in the case of bosonic spinon and fermonic holon can be derived and solved in a straightforward way using the approach discussed here by changing asymmetry angles $\theta_f\to \theta_{\fb}$ and $\theta_b\to \theta_{\ff}$.

The saddle point equations of interest at zero temperature are those in \eqref{Eq:EoM1}-\eqref{Eq:EoM4}.
To solve these equations numerically, we use an approach similar to the one used in the previous paper \cite{Tikhanovskaya:2020elb}. The detailed treatment of the equations is discussed in the Appendix \ref{app:tJ-T0}. Our focus is to solve the saddle point equations for bosonic and fermionic spectral densities first, and then present results for gauge invariant observables such as electron and spin spectral densities, and optical conductivity. 

We focus on the asymmetry parameters of the model $\theta_f$, $\theta_b$ that satisfy the Luttinger constraints. In this section we are interested in a particular case of $\Delta_b=\Delta_f=1/4$, and thus we assume it everywhere. Therefore, we can write the Luttinger constraints (\ref{eq:LtB}) in the following form
\begin{eqnarray}
  \frac{\theta_f}{\pi}+\frac14{\sin(2\theta_f)} &=& \frac{1}{2}-\kappa+kp, \label{eq:Ltfdel14}\\
  \frac{\theta_b}{\pi}+\frac14{\sin(2\theta_b)} &=& \frac{1}{2}+p\,. \label{eq:Ltbdel14}
\end{eqnarray}
The solutions with the constrains will correspond to the region crossed by one of the lines at chosen $\kappa$ in the Fig.~\ref{fig:diagram_fermionic_spin}. Although there is a large region of asymmetry parameters where an analytical solution approximates a numerical result well, in this section we analyze only a portion of the full parameter space. We choose the asymmetry angles in the region of $\theta_b=0.26\pi,0.32\pi$ and $0<\theta_f\leq0.15\pi$.

\begin{figure}[h!]
(a)\includegraphics[width=0.447\textwidth]{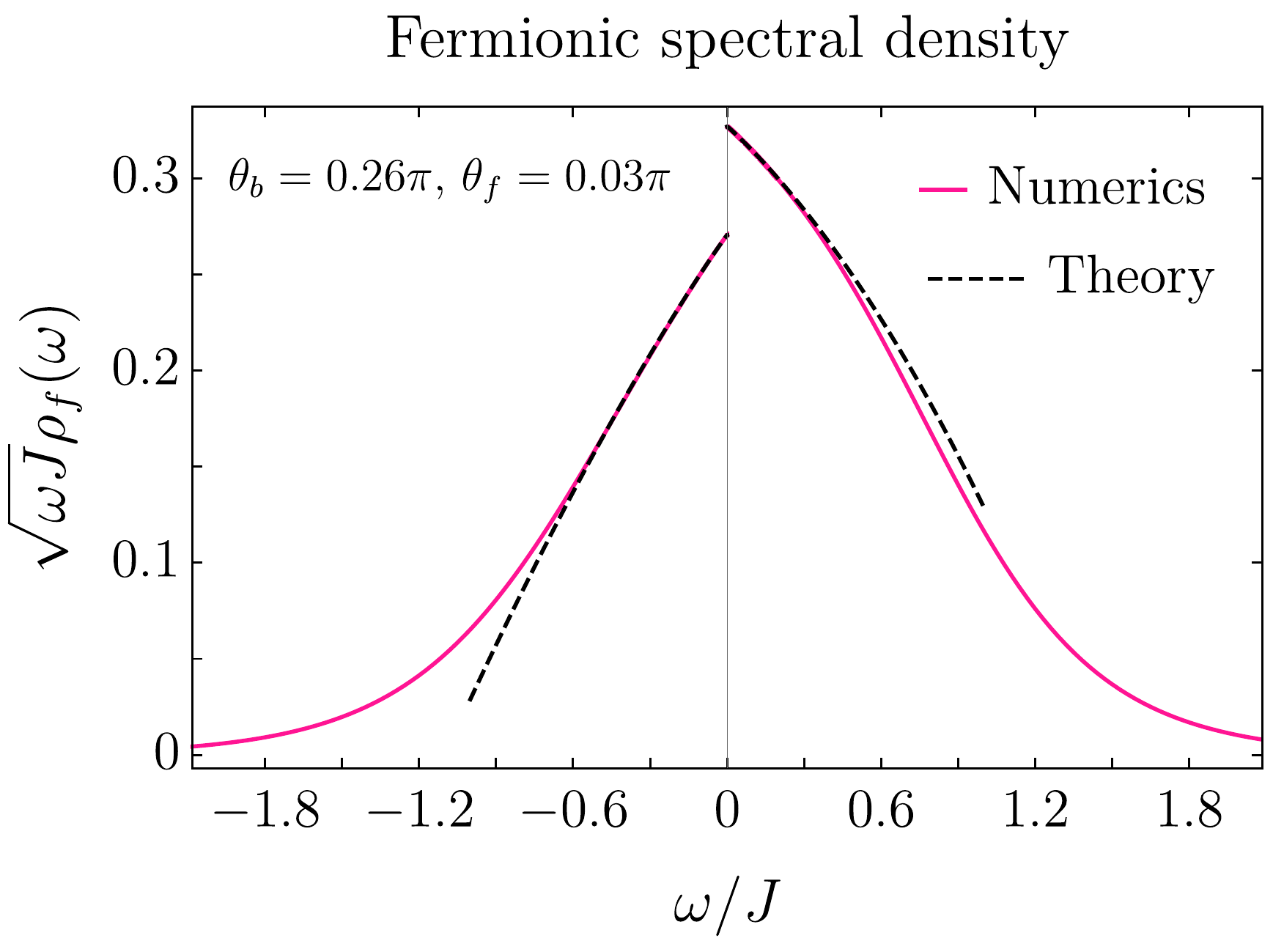}\quad\quad
(b)\includegraphics[width=0.447\textwidth]{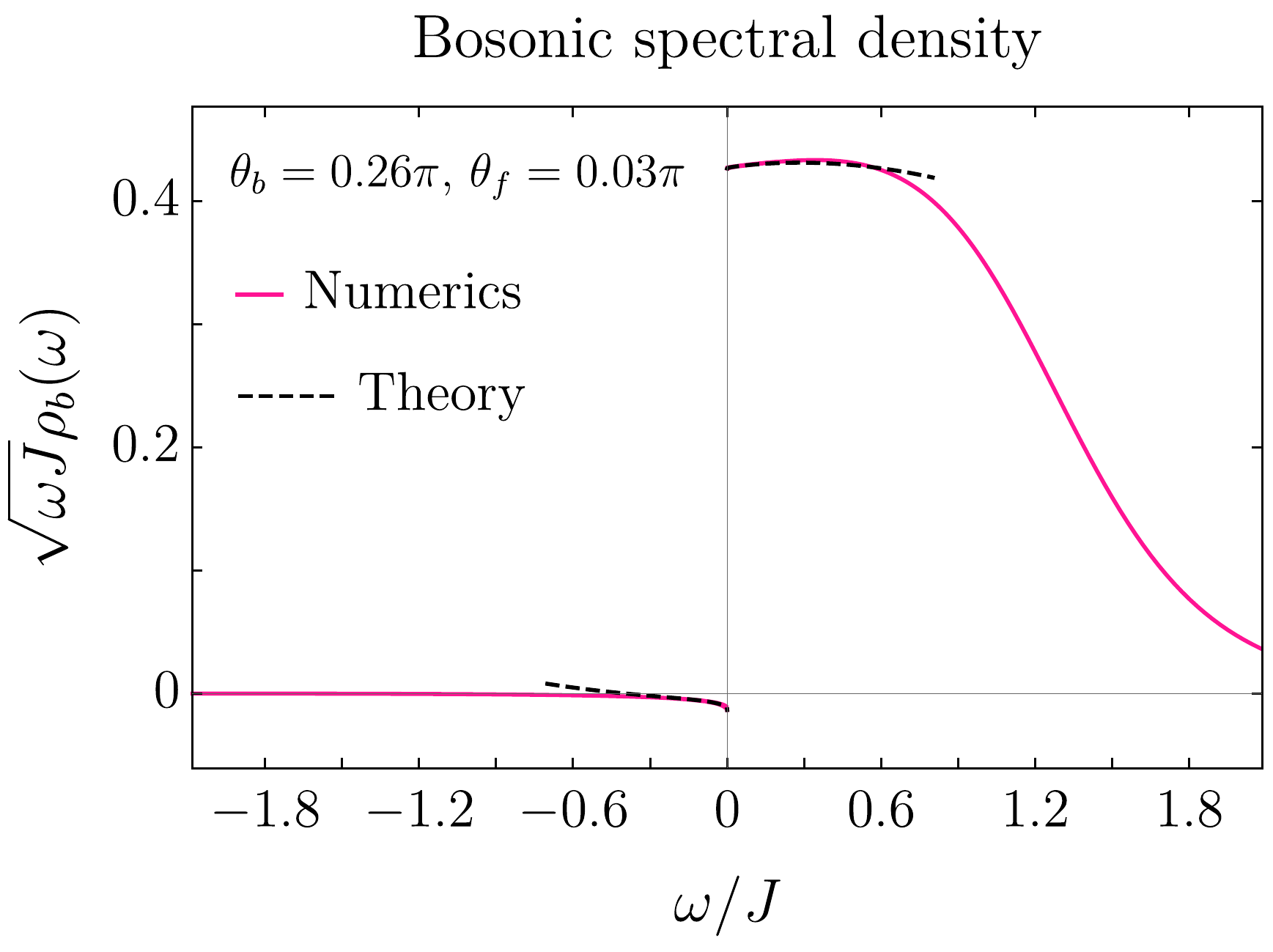}
\caption{Fermioic (a) and bosonic (b) spectral densities as solutions of the equations \eqref{Eq:EoM1}-\eqref{Eq:EoM4}. In both plots the chosen parameters: $\theta_b=0.26\pi$, $\theta_f=0.03\pi$. Solid magenta lines are the numerical solution for these parameters. Black dashed lines are given by analytical solution with nonlinear corrections. The values of the fitting parameters are given in the Fig.~\ref{table:alpha_h}. \label{Fig:NumerSpd}}
\end{figure}

\begin{figure}[h!]
(a)\includegraphics[width=0.447\textwidth]{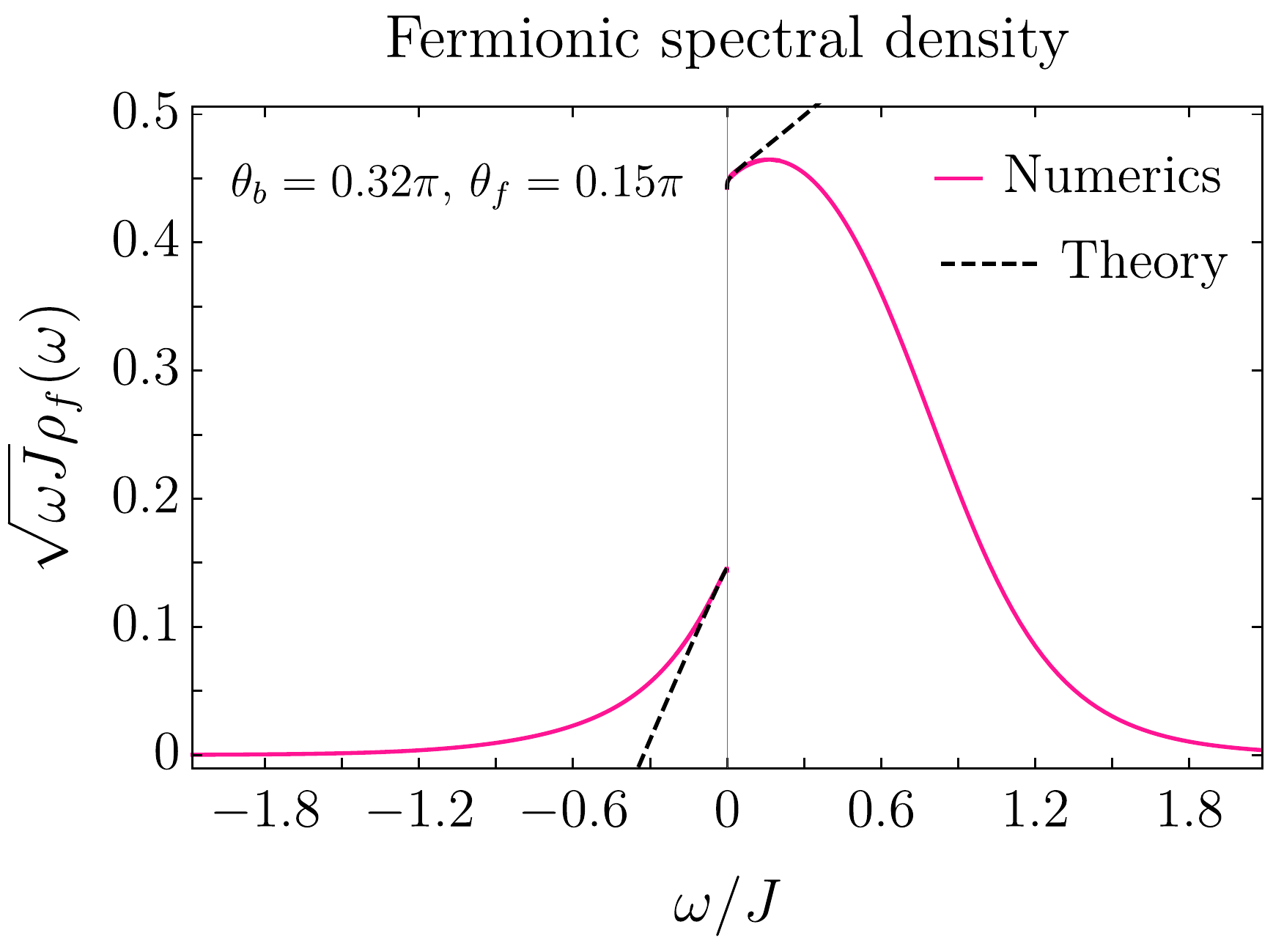}\quad\quad
(b)\includegraphics[width=0.447\textwidth]{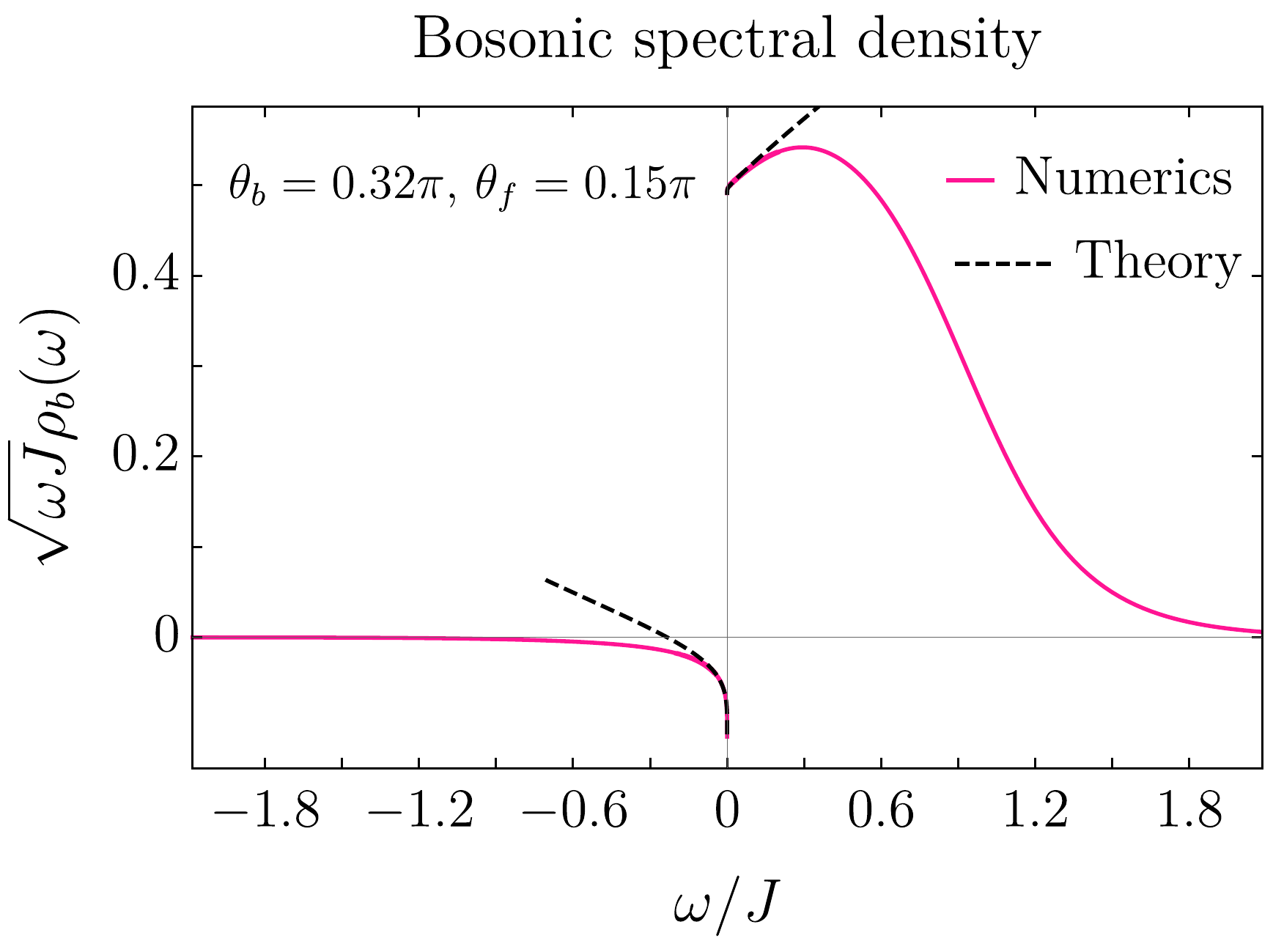}
\caption{Fermioic (a) and bosonic (b) spectral densities as solutions of the equations \eqref{Eq:EoM1}-\eqref{Eq:EoM4}. As in the Fig.~\ref{Fig:NumerSpd} except the chosen parameters are $\theta_b=0.32\pi$, $\theta_f=0.15\pi$.\label{Fig:NumerSpd15}}
\end{figure}

We now show that the numerical solution with the constraints implied in small enough $\omega$-region has behavior given by the results of the theoretical approach in the previous section. We discuss the behavior of fermionic and bosonic spectral densities as solutions of the saddle point equations \eqref{Eq:EoM1}-\eqref{Eq:EoM4} as well as gauge invariant observables such as electron, spin spectral densities and optical conductivity at zero temperature.

\begin{figure}[h!]
\includegraphics[width=0.4\textwidth]{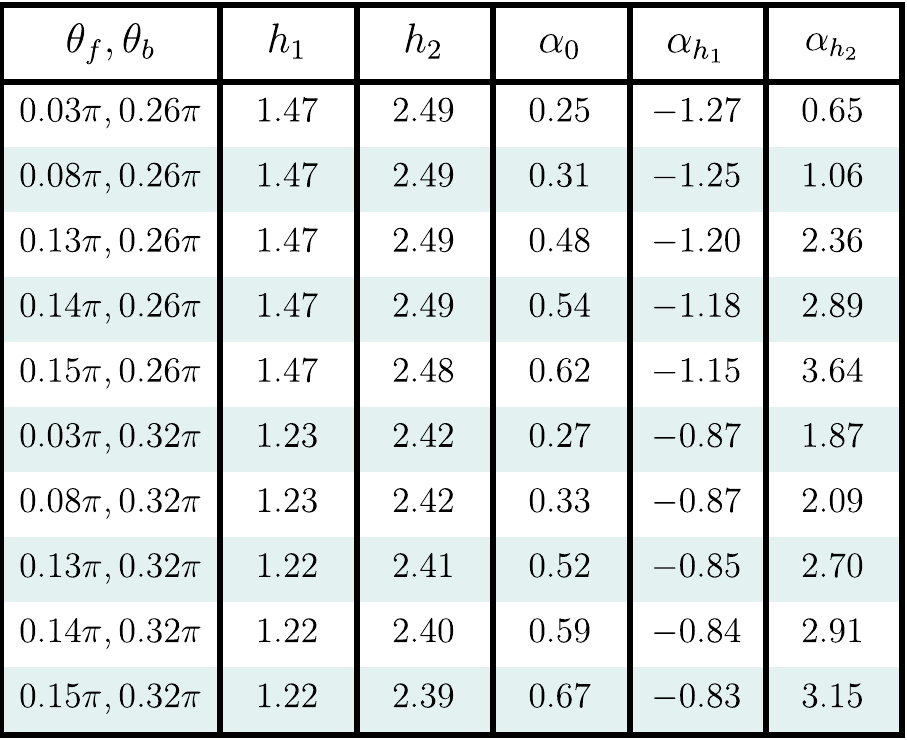}
\caption{Approximate values of exponents $h$ and coefficients $\alpha_h$ for different asymmetry angles $\theta_f$, $\theta_b$ obtain by fits to the numerical solution. \label{table:alpha_h}}
\end{figure}

We first find the solution for bosonic $\rho_b(\omega)$ and fermionic $\rho_f(\omega)$ spectral densities of the equations \eqref{eq:Ltbdel14}-\eqref{eq:Ltbdel14} at zero temperature at various sets of parameters $\theta_f$ and $\theta_b$ and fixed $k=0.2$, $t=J=1$. The method of solving the equations at zero temperature is explained in detail in Appendix \ref{app:tJ-T0} as well as in \cite{Tikhanovskaya:2020elb}. Amongst the range of parameters for which we solve the equations, we show two solutions that represent the behavior of the spectral densities at large ($\theta_b=0.32\pi$, $\theta_f=0.15\pi$) and small ($\theta_b=0.26\pi$, $\theta_f=0.03\pi$) asymmetry angles. The solutions are presented in Figs.~\ref{Fig:NumerSpd},\ref{Fig:NumerSpd15}, respectively.

The solutions found in this section compared to an analytical fit at small frequencies described in the previous sections and in Ref.~\onlinecite{Tikhanovskaya:2020elb}. Using the numerical solutions, we find the threshold of asymmetry angles where it is still possible to get the solution as discussed in Appendix \ref{app:tJ-T0}. Below we describe a behavior of the gauge invariant observables and show that an analytical solution follows a numerical curve well enough at small frequencies.

Let us first rewrite the analytical expressions of the polynomials from section \ref{sec:GIO} that, as we show below, approximate the exact numerical solution at small frequencies. Following the notations in the previous sections, we use $h_1$, $h_2$ to denote the exponents in the fitting functions. We consider the polynomials for each of the gauge invariant observables truncated at the order of $h_2-1$ in frequency with all non-linear corrections included. A truncated expression for electron and spin spectral densities \eqref{eq:rhocT=0}, \eqref{eq:rhoQT=0} can be written as
\begin{equation}\label{eq:FitEl}
\begin{split}
\rho_{e,Q}(\omega)=\rho_{e,Q}^c(\omega)\left(1- \sum_{h}^{h_2}\alpha_{h}A^{e,Q}_{h}|\omega/{J}|^{h-1}-\sum_{hh'}^{h_2}\alpha_h\alpha_{h'}A^{e,Q}_{hh'}|\omega/{J}|^{h+h'-2}\right).
 \end{split}
\end{equation}
We write the similar expression for optical conductivity $\sigma(\omega)$ truncating the full series \eqref{eq:sigmaT=0} in a similar way
\begin{equation}\label{eq:FitCond}
\sigma(\omega)=\sigma_0(\omega)\left(1- \sum_{h}^{h_2}\alpha_{h}A^\sigma_{h}|\omega/{J}|^{h-1}-\sum_{hh'}^{h_2}\alpha_h\alpha_{h'}A^\sigma_{hh'}|\omega/{J}|^{h+h'-2}\right)
\end{equation}
with different coefficients $A^\sigma_h$. In the above, $\rho_{e,Q}^c(\omega)$ are the electron and spin spectral weights of the conformal solutions \eqref{eq:rhocConf}, \eqref{eq:Qconf}, and $\sigma_0$ is given by \eqref{eq:sigma0T0}. The coefficients $A^{e,Q,\sigma}_h$ are known exactly and have analytical forms for electron spectral density \eqref{eq:rhocT=0}, for spin spectral density \eqref{eq:rhoQT=0} and  for conductivity \eqref{eq:sigmaT=0}.
\begin{figure}[h!]
\includegraphics[width=.8\textwidth]{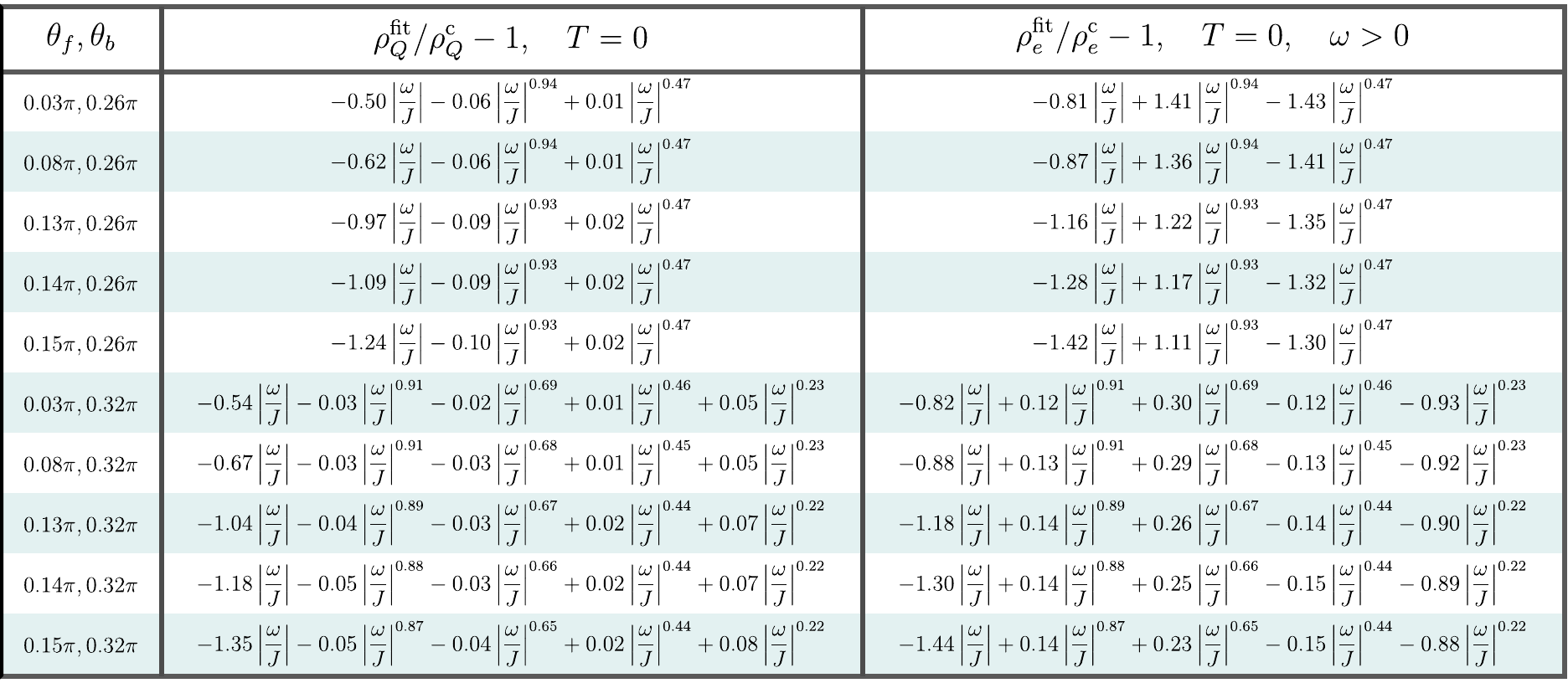}
\caption{Table of computed analytical expressions of spin and electron spectral densities \eqref{eq:FitEl} with all parameters fixed for a range of asymmetry angles $(\theta_f,\theta_b)$. The polynomials approximate a corresponding numerical solution at small frequencies. The exponents are given by $h-1$ where $h$ are presented in the Fig.~\ref{table:alpha_h}.  \label{table:rhoeQ}}
\end{figure}

\begin{figure}[h!]
(a)\includegraphics[width=0.447\textwidth]{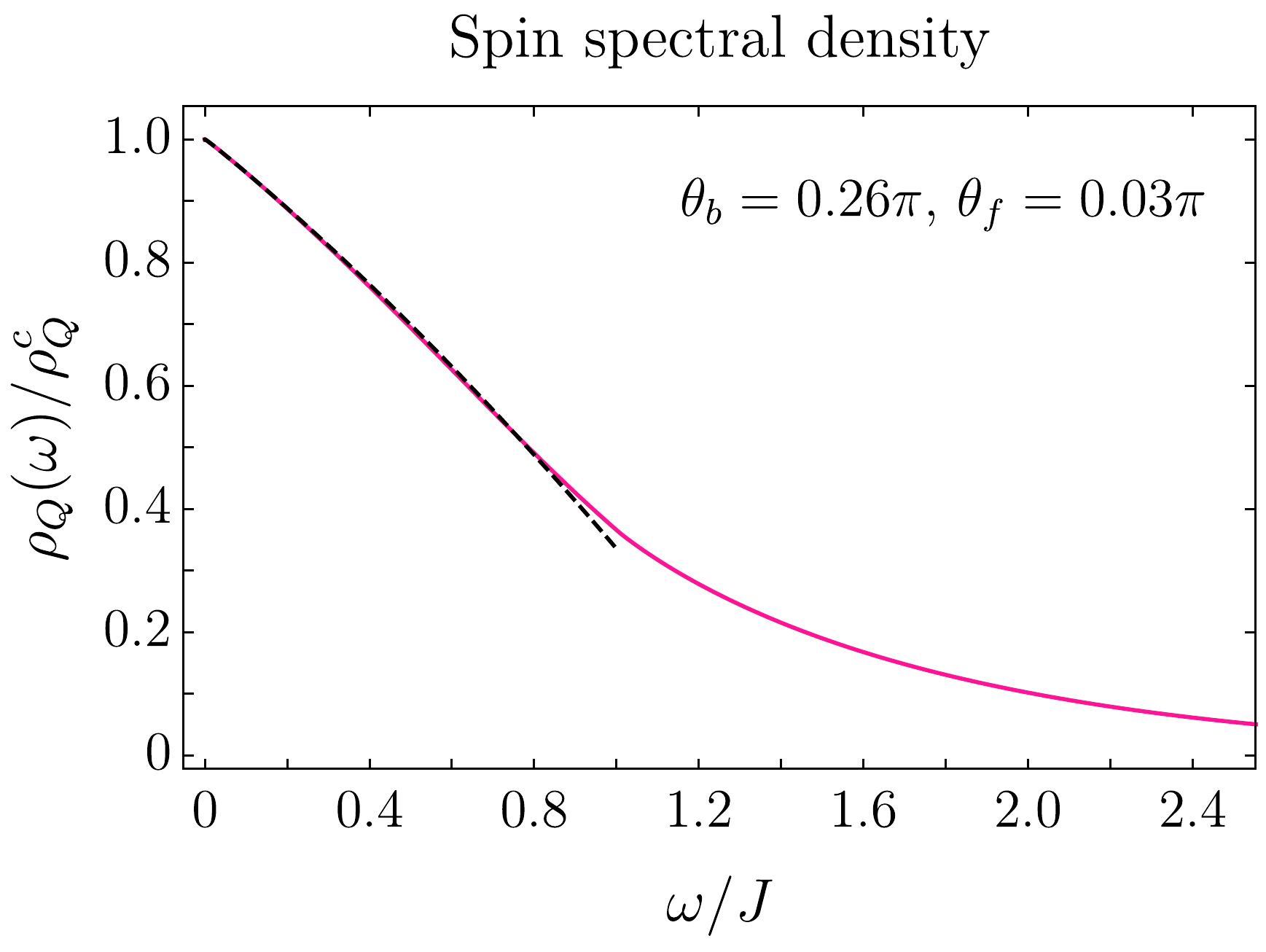}\quad\quad
(b)\includegraphics[width=0.447\textwidth]{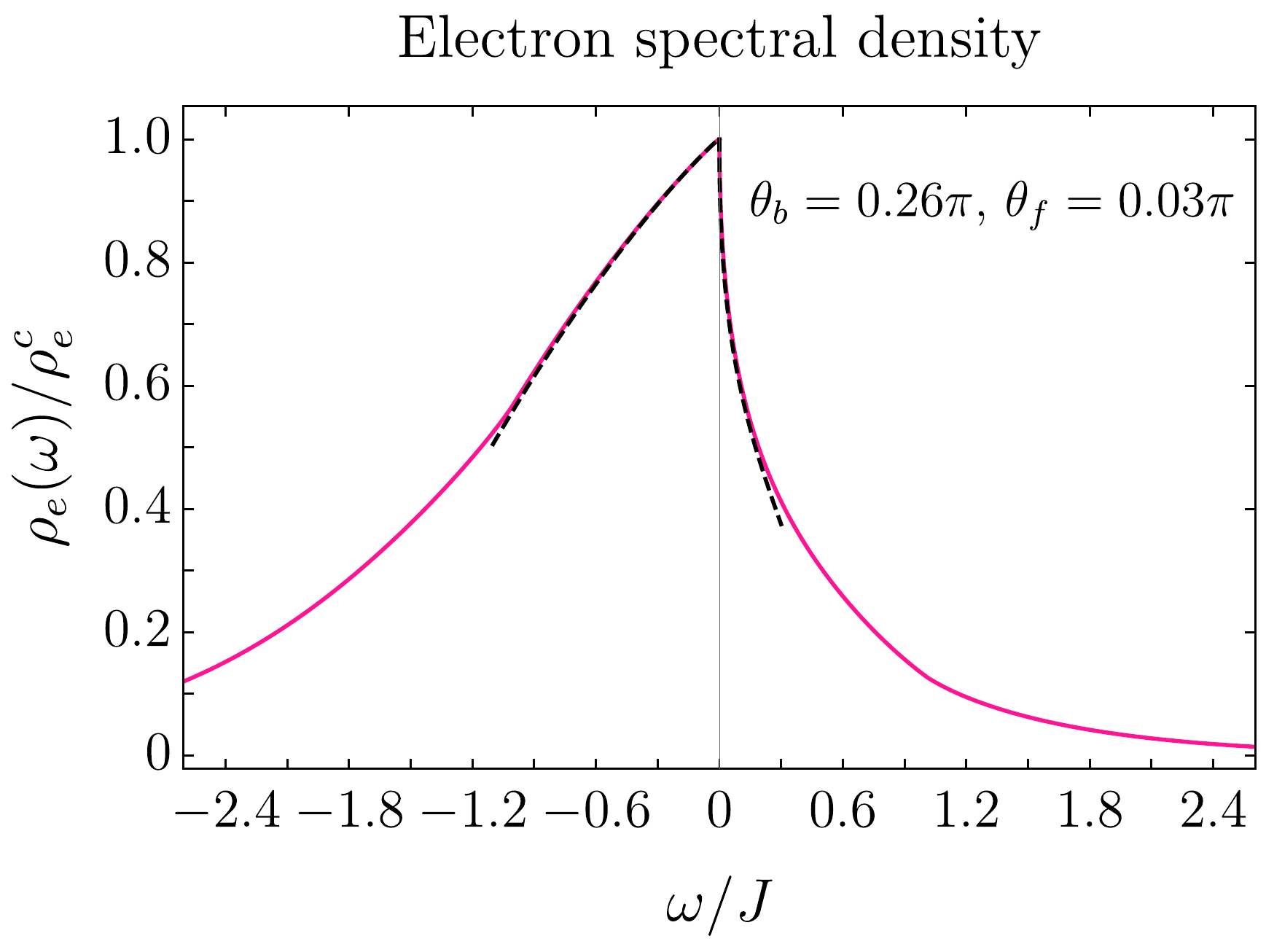}
\caption{Spin (a) and electron (b) spectral densities. Solid magenta lines are numerical solutions. Black dashed lines are given by analytical solution \eqref{eq:FitEl}. Here $\theta_b=0.26\pi$ and $\theta_f=0.03\pi$. The exponents and fitting functions are given in Figs.~\ref{table:alpha_h}, \ref{table:rhoeQ} respectively. \label{Fig:NumerSpin}}
\end{figure}

\begin{figure}[h!]
(a)\includegraphics[width=0.447\textwidth]{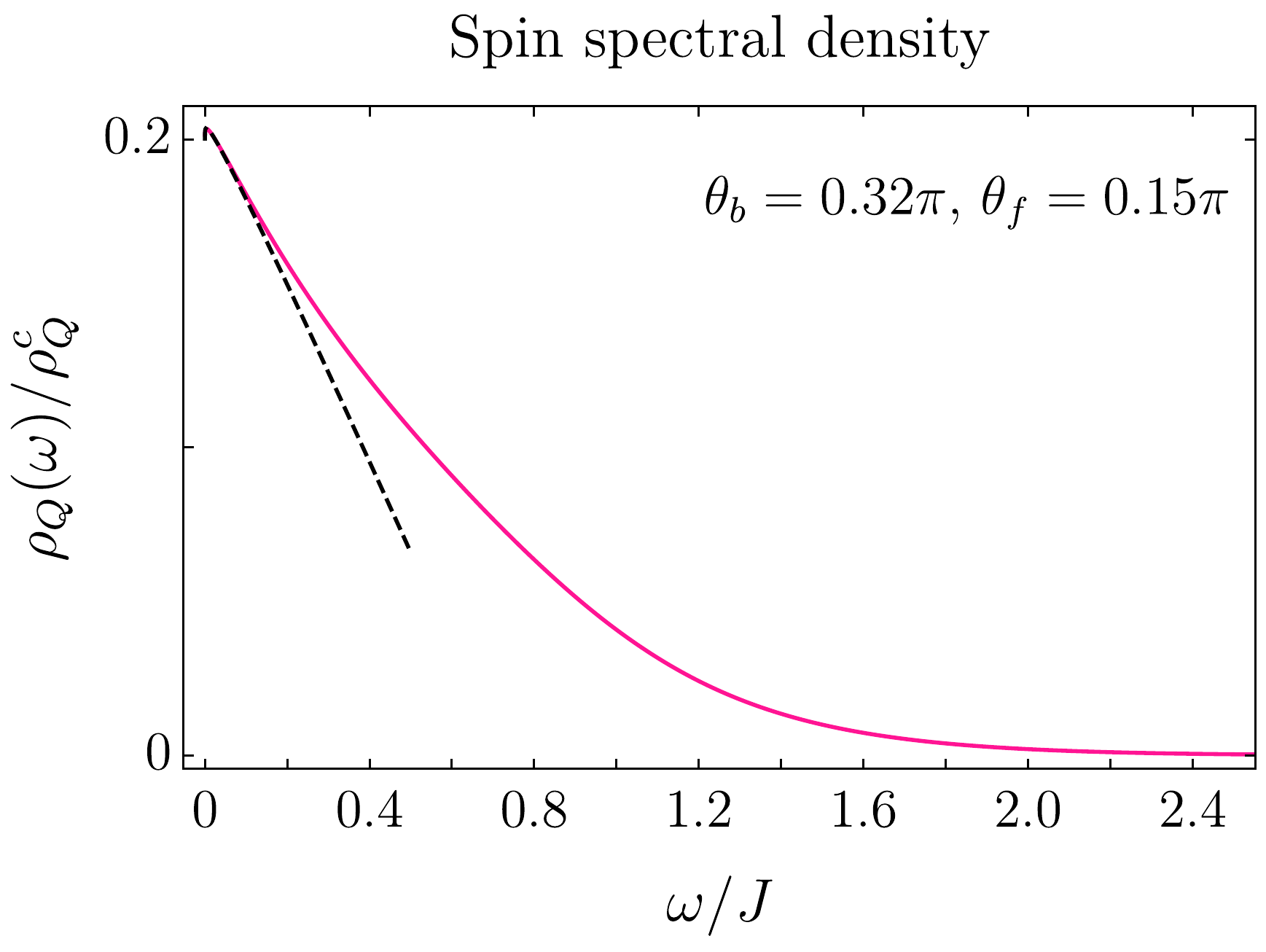}\quad\quad
(b)\includegraphics[width=0.447\textwidth]{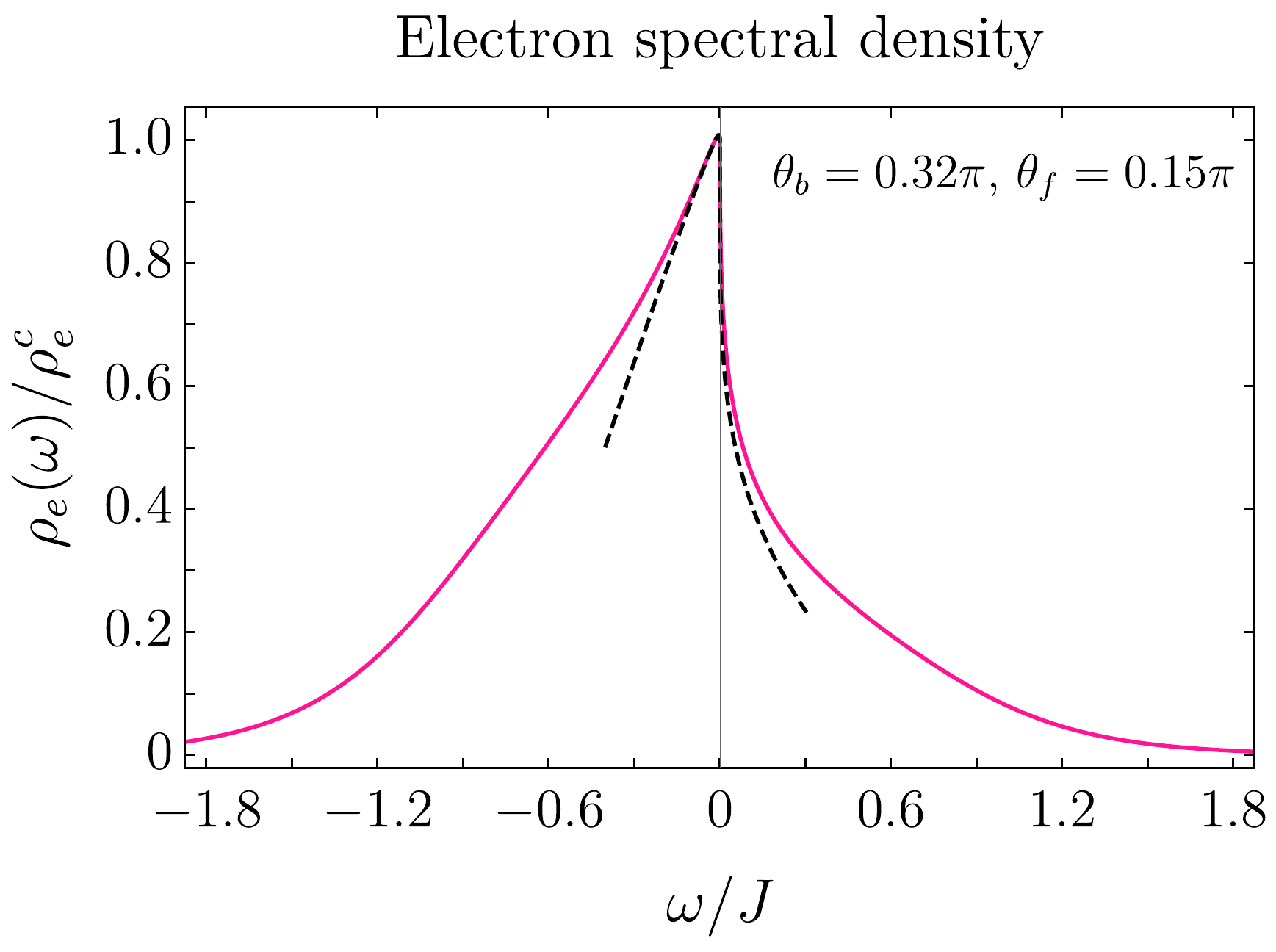}
\caption{Spin (a) and electron (b) spectral densities. Same as in the Fig.\ref{Fig:NumerSpin} except the parameters $\theta_b=0.32\pi$ and $\theta_f=0.15\pi$. \label{Fig:NumerSpin15}}
\end{figure}

We note that the analytical solution has unfixed parameters $\alpha_h$. In the series we consider, there are three parameters the values of which can be obtained by fitting an analytical expression into to a numerical curve at corresponding parameters. For a range of parameters we discuss in this section, the values of coefficients $\alpha_h$ as well as exponents $h$ are presented in Fig.~\ref{table:alpha_h}.

\begin{figure}[h!]
\includegraphics[width=.6\textwidth,height=0.4\textwidth]{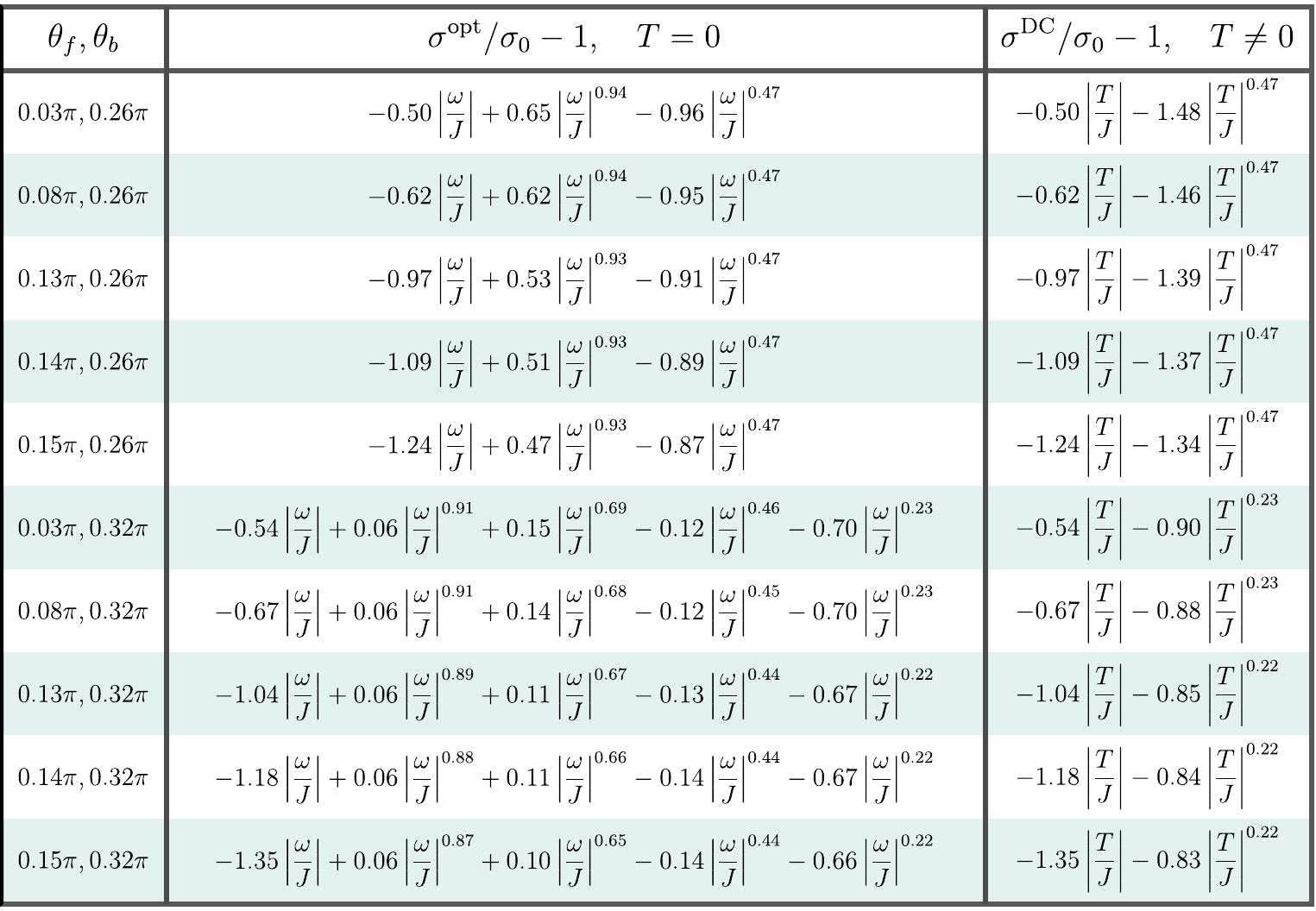}
\caption{Table of computed analytical expressions of optical conductivity \eqref{eq:FitCond} at zero temperature (second column) and DC conductivity \eqref{eq:sigmaDCT} at finite temperature with all parameters fixed for a range of asymmetry angles $(\theta_f,\theta_b)$. The polynomials approximate a corresponding numerical solution at small frequencies. The exponents are given by $h-1$ where the values of $h$ are presented in the Fig.~\ref{table:alpha_h}.  \label{table:sigmaT}}
\end{figure}

Using the values of the coefficients $\alpha_h$ we obtain analytical solutions with all parameters fixed. The analytical formulas for spin and electron spectral densities are presented in the Figs.~\ref{table:rhoeQ}, \ref{table:sigmaT}. From the polynomials in the tables we notice that increasing the asymmetry angles leads to an increasing a linear coefficient compared to the coefficient of the $h_1-1$ exponent and coefficients of its non-linear corrections. By making it big enough, the linear coefficient becomes significantly larger than other lower order coefficients. We show two examples of fitting the corresponding analytical expressions of the spin and electron spectral densities in the Figs.~\ref{Fig:NumerSpin},\ref{Fig:NumerSpin15} for different asymmetry angles.

\begin{figure}[h!]
(a)\includegraphics[width=0.447\textwidth]{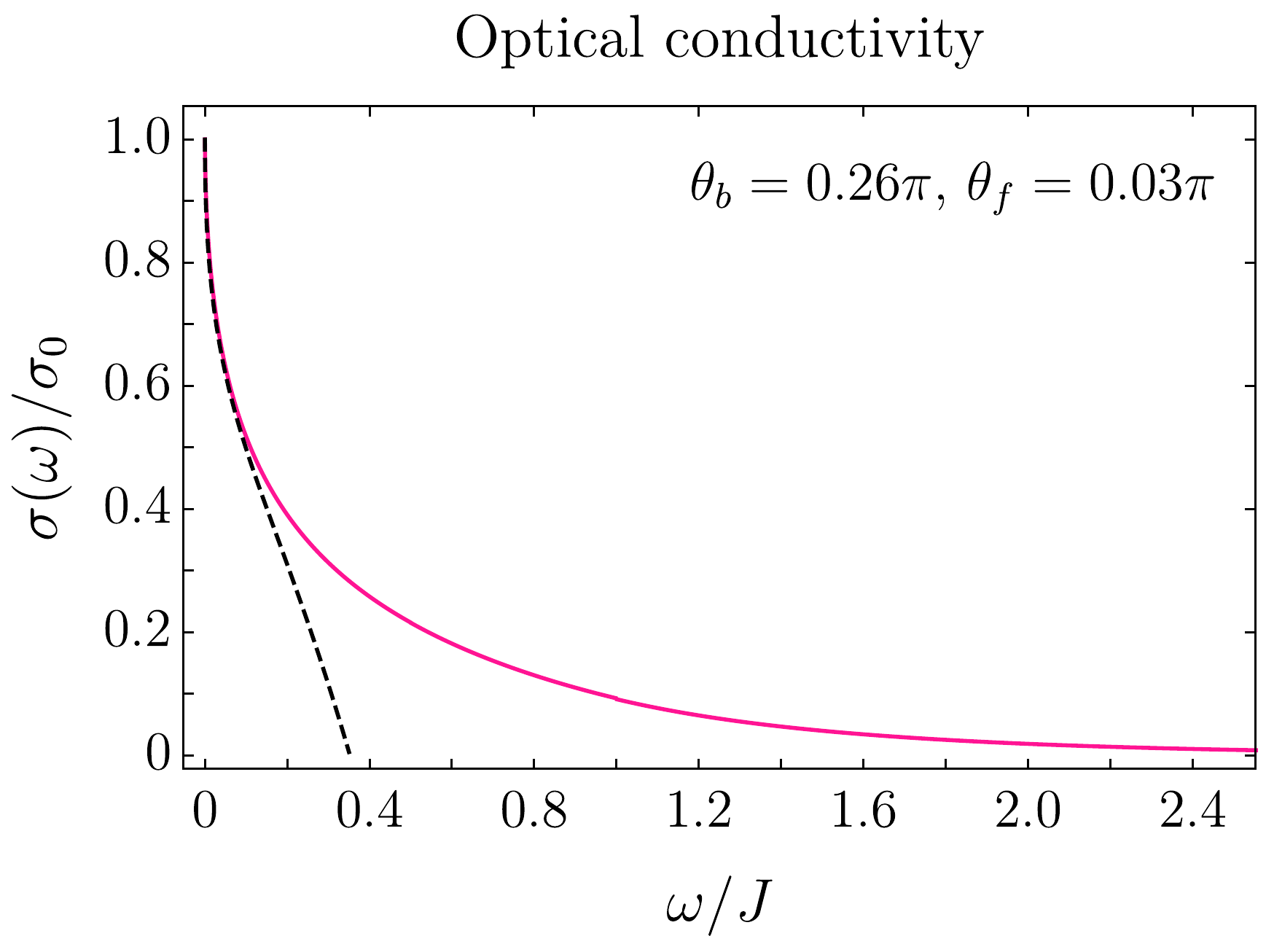}\quad
(b)\includegraphics[width=0.447\textwidth]{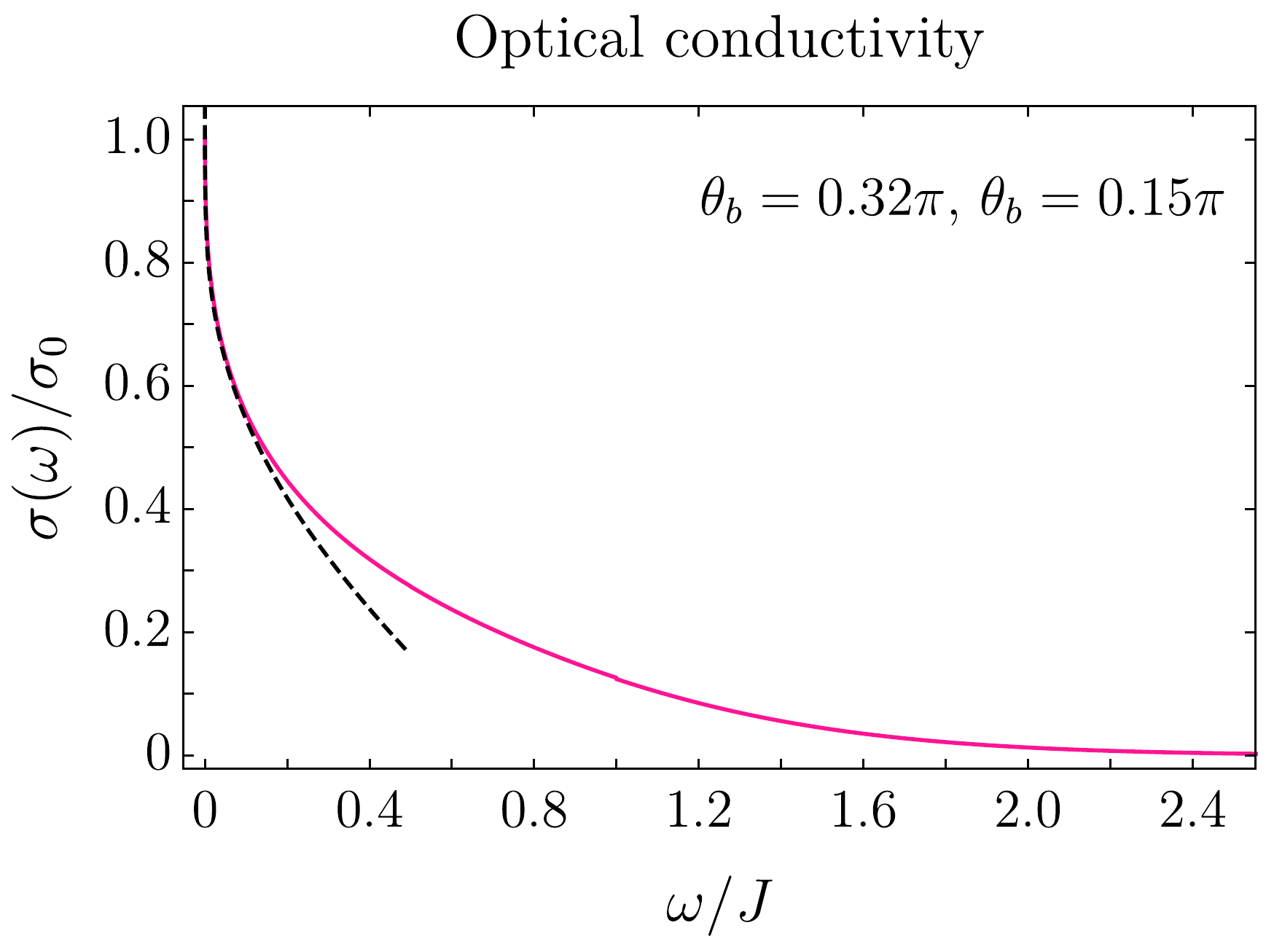}
\caption{Optical conductivity for (a) $\theta_b=0.26\pi$ and $\theta_f=0.03\pi$ and (b) $\theta_b=0.32\pi$ and $\theta_f=0.15\pi$. Numerical solutions in both plots are given by the solid magenta lines and the black dashed lines are the analytical solutions presented in Fig.~\ref{table:sigmaT}. \label{Fig:cond}}
\end{figure}

We conclude this section by discussing the optical and d.c. conductivity. In the Table~\ref{table:sigmaT} we present the analytical expressions for optical conductivity at zero temperature and finite temperature d.c. conductivity with all parameters fixed. We note that even though we do not discuss the finite temperature exact numerical solution in this paper, the coefficients $\alpha_h$ are fixed to their zero temperature values, and therefore, we can write down the analytical formulas at finite temperature as well. We again notice increasing the asymmetry angles leads to increasing the linear coefficient. Within the allowed range of asymmetry angles ($\theta_f\leq 0.15\pi$, as discussed in the Appendix \ref{app:tJ-T0}) we see that the linear term becomes dominant. We explicitly show a behavior of optical conductivity at zero temperature in Fig.~\ref{Fig:cond} for different asymmetry angles.

\section{Conclusions}
\label{sec:conc}

We have presented numerical and analytic solutions of the non-Fermi liquid phase random $t$-$J$ model without quasiparticle excitations, with good agreement between the two approaches. The solutions were obtained the large $M$ limit of a model with SU($M$) spin symmetry, and follow our previous analyses of the undoped insulator in Ref.~\onlinecite{Tikhanovskaya:2020elb}. The theory has the structure of the SYK model for fractionalized spinons and holons carrying gauge charges, which have random $q=4$-body interaction terms between them. Note, however, that expressed in terms of the underlying gauge-invariant electrons, the interactions are not of the $q=4$ SYK form.

Our attention focused on the leading and sub-leading singularities of several gauge-invariant physical observables at low frequencies and temperatures. These are controlled by operators with scaling dimension $h$ which perturb the leading conformally invariant solution. As the theory is in $0+1$ spacetime dimensions, we require $h>1$ for the operator to be irrelevant in the infrared, and contribute to the sub-leading singularities.

An important operator that appears in all the subleading corrections is the time-reparameterization mode, or the boundary graviton of the holographic dual. This operator is universal in SYK criticality, and has $h_0=2$ exactly; no corrections are expected to this scaling dimension at all orders in $1/M$. It was also present in the previous analysis of the insulating model \cite{Tikhanovskaya:2020elb}. This operator leads to relative corrections in observables which scale $\omega^{h_0-1}$ or $T^{h_0-1}$ in the IR. In particular, as the leading contribution to the resistivity is $T$-independent in the $t$-$J$, the first contribution from the graviton mode is a linear-in-$T$ resistivity \cite{Guo:2020aog}.

We also found that the $t$-$J$ model has an irrelevant operator with scaling dimension $1 < h_1 < 2$ over the regime where scaling solutions were found numerically (see Fig.~\ref{table:alpha_h}). This will therefore lead to corrections which scale as $\omega^{h_1-1}$ or $T^{h_1-1}$ in the IR, which are more singular than those from the graviton mode. This operator is special to the doped case, and was not found in insulator studied previously \cite{Tikhanovskaya:2020elb}.
However, it should be noted that this operator is not protected, and we do expect $1/M$ corrections to the value of $h_1$. We have not studied these corrections, and the possibility remains that such corrections will increase the value of $h_1$ above 2, and return the graviton mode as the dominant correction. Moreover, we found that the numerical co-efficient of the $h_1$ correction was quite small in the spin spectral density, and significantly smaller than that of the $h_0$ correction (see Fig.~\ref{table:rhoeQ}).

A recent exact-diagonalization study of the random $t$-$J$ model with SU(2) spin symmetry \cite{Shackleton20} computed the spin spectral density and found that it matched the spectral density similar to Fig.~\ref{Fig:NumerSpin}a over a wide range of frequencies near the critical doping where the spin-glass order vanished (spin-glass order is present at $M=2$ for low doping). The numerics show $\rho_Q (\omega) \sim \mbox{sgn}(\omega) \left[c_0 - c_1 |\omega| + \ldots \right]$ at small $\omega$, consistent with the numerical results in Table~\ref{table:rhoeQ} where the $h_1$ contribution is small.

We note from Figs.~\ref{table:rhoeQ} and~\ref{table:sigmaT} that the contributions of $h_1$ mode are more significant for the other observables, including the d.c. resistivity, and can be comparable to the $h_0=2$ contribution. It would be interesting to study the relative contributions of these modes at higher order in $1/M$ to understand the situation for SU(2) better.

\subsection*{Acknowledgements}

We thank Yingfei Gu, Igor R. Klebanov, and Henry Shackleton for valuable discussions.
This research was supported by the U.S. Department of Energy under Grant DE-SC0019030.

\appendix

\section{Zero temperature numerics for $t-J$ model}\label{app:tJ-T0}
We consider Dyson-Schwinger equations for the retarded Green's function for $t-J$ model, which is obtained by analytic continuation from the Matsubara frequency $i\omega_{n}\to \omega +i 0$. The first pair of the Dyson-Schwinger  equations reads
\begin{align}\label{firstDS-ferm}
   & G_{f,R}(\omega)^{-1} = \omega +i0 +\mu_f-\Sigma_{f,R}(\omega)\,;\\\label{firstDS-bos}
   & G_{b,R}(\omega)^{-1} = \omega  +i0 +\mu_b-\Sigma_{b,R}(\omega)\,.
\end{align}

As before, we define analytic in the upper half plane Green's functions $G(z)$ and self energy $\Sigma(z)$, which are expressed through
the spectral densities $\rho(\omega)$ and $\sigma(\omega)$ as
\begin{align}
G(z) =  \int_{-\infty}^{+\infty} d\omega\frac{\rho(\omega)}{z-\omega}\,, \quad
\Sigma(z) =  \int_{-\infty}^{+\infty} d\omega\frac{\sigma(\omega)}{z-\omega}\,. \label{spdGSigma}
\end{align}
The Matsubara and retarded Green's functions can be obtained from these functions by taking $z=i\omega_{n}$ and $z=\omega+i0$. We can find the spectral density as $\rho(\omega)= -\frac{1}{\pi}\textrm{Im}G_{R}(\omega)$. Also using the representation (\ref{spdGSigma}) we can obtain Green's function in imaginary time expressed through integral over the spectral density
\begin{align}
G_a(\tau)=\frac{1}{\beta}\sum_{n}G_a(i\omega_{n})e^{-i\omega_{n}\tau}=-\int_{-\infty}^{+\infty}d\omega \frac{\rho(\omega) e^{-\omega \tau}}{1-\zeta e^{-\beta \omega}}, \quad \tau \in(0,\beta)\,, \label{GtaurhoTJ}
\end{align}
where $a=f,b$ and $\zeta_f = -1$, $\zeta_b = 1$. We  notice that $G(0^{+})-\zeta G(\beta^{-})=-1=-\int_{-\infty}^{+\infty}d\omega \rho(\omega)$ for arbitrary temperature.


Here we consider an original model with fermionic spinon and bosonic holon, that gives the following second pair of the Dyson-Schwinger equations
\begin{align}
    &\Sigma_f(\tau) = J^2 G_f^2(\tau)G_f(\beta-\tau)  + kt^2 G_f(\tau)G_b(\tau)G_b(\beta-\tau)\,;\\
    &\Sigma_b(\tau) = t^2 G_f(\tau)G_f(\beta -\tau)G_b(\tau)\,.
\end{align}

To obtain the second Dyson-Schwinger equation for the retarded self-energy $\Sigma_{R}(\omega )$ we consider the equations in the Matsubara  space above and use (\ref{GtaurhoTJ}) to write it in terms of the spectral densities for fermions
\begin{align}
\Sigma_f(i\omega_{n}) &=-J^{2}\int_{-\infty}^{+\infty}\prod_{i=1}^{3}(d\omega_{i} \rho_f(\omega_{i})) \frac{n_{f}(\omega_{1})n_{f}(\omega_{2})n_{f}(-\omega_{3})+n_{f}(-\omega_{1})n_{f}(-\omega_{2})n_{f}(\omega_{3})}{\omega_{1}+\omega_{2}-\omega_{3}-i\omega_{n}}\\
& + kt^2 \int_{-\infty}^{+\infty}\prod_{i=1}^{3}d\omega_{i} \rho_f(\omega_{1})\rho_b(\omega_{2})\rho_b(\omega_{3}) \frac{n_{f}(\omega_{1})n_{b}(\omega_{2})n_{b}(-\omega_{3})+n_{f}(-\omega_{1})n_{b}(-\omega_{2})n_{b}(\omega_{3})}{\omega_{1}+\omega_{2}-\omega_{3}-i\omega_{n}}
\end{align}
and for bosons
\begin{align}
\Sigma_b(i\omega_{n}) & =t^2 \int_{-\infty}^{+\infty}\prod_{i=1}^{3}d\omega_{i} \rho_f(\omega_{1})\rho_b(\omega_{2})\rho_f(\omega_{3}) \frac{n_{f}(\omega_{1})n_{b}(\omega_{2})n_{f}(-\omega_{3})+n_{f}(-\omega_{1})n_{b}(-\omega_{2})n_{f}(\omega_{3})}{\omega_{1}-\omega_{2}+\omega_{3}-i\omega_{n}}
\end{align}
where $n_{a}(\omega)=1/(e^{\beta \omega}-\zeta)$ is the Bose or Fermi distribution and we can get $
\Sigma_{R}(\omega ) =\Sigma(i\omega_{n}=\omega+i0)$. At zero temperature $\beta=\infty$ we can replace $n_{b}(\omega)$ by $-\theta(-\omega)$ and $n_{f}(\omega)$ by $\theta(-\omega)$. Though $n_{b}(\omega)$ is divergent for $\omega \to 0$, we assume that this divergence does not play any role.
Functions $G_{R}(\omega)$ and $\Sigma_{R}(\omega)$ are complex valued and further we will adopt notations for their real and imaginary parts $G_{R}(\omega) =G'(\omega)+iG''(\omega)$ and $\Sigma_{R}(\omega)=\Sigma'(\omega)+i\Sigma''(\omega)$. So for $\beta=\infty$ for fermions we find
\begin{align}
\Sigma_f''(\omega) = \begin{cases}
\pi \int_{0}^{\omega_{1}+\omega_{2}\leq \omega}d\omega_{1}d\omega_{2}\rho_f(\omega_{1})\left(kt^2\rho_b(\omega_{2})\rho_b(\omega_{1}+\omega_{2}-\omega) - J^2\rho_f(\omega_2)\rho_f(\omega_1+\omega_2 -\omega)\right),\quad \omega >0\\
\pi \int^{0}_{\omega_{1}+\omega_{2} \geq \omega}d\omega_{1}d\omega_{2}\rho_f(\omega_{1})\left(kt^2\rho_b(\omega_{2})\rho_b(\omega_{1}+\omega_{2}-\omega) - J^2\rho_f(\omega_2)\rho_f(\omega_1+\omega_2 -\omega)\right), \quad \omega <0\, \label{sigmadpfermTJ}
\end{cases}
\end{align}
and for bosons
\begin{align}
\Sigma_b''(\omega) = \begin{cases}
-\pi t^2 \int_{0}^{\omega_1 + \omega_2 \leq \omega} d\omega_{1}d\omega_{2}\rho_f(\omega_{1})\rho_b(\omega_{2})\rho_f(\omega_{2}+\omega_{1}-\omega),\quad \omega >0\\
-\pi t^2  \int_{\omega_1 + \omega_2 \geq \omega}^{0} d\omega_{1}d\omega_{2}\rho_f(\omega_{1})\rho_b(\omega_{2})\rho_f(\omega_{2}+\omega_{1}-\omega), \quad \omega <0\,. \label{sigmadpbosTJ}
\end{cases}
\end{align}
We anticipate that at zero temperature the functions $\rho(\omega)$ and $\Sigma''(\omega)$ will have discontinuity. So it will be convenient to use a new set of functions defined separately for $\omega >0$ and $\omega <0$
\begin{align}
\rho_a(\omega)=
\begin{cases}
\frac{g_{a}^+(\omega)}{\sqrt{\omega}}, \quad \omega >0\\
\frac{g_{a}^-(-\omega)}{\sqrt{-\omega}}, \quad \omega <0
\end{cases}\,, \quad
\Sigma_a''(\omega) = \begin{cases}
4\pi\sqrt{\omega} s_{a}^+(\omega), \quad \omega>0 \\
4\pi\sqrt{-\omega} s_{a}^-(-\omega), \quad \omega<0
\end{cases}\,. \label{defrhoSigmaImTJ}
\end{align}
We make change of variables $\omega_{1}=\omega \sin^{2} u \cos^{2}\phi$ and $\omega_{2}=\omega \sin^{2} u \sin^{2}\phi$ in (\ref{sigmadpfermTJ}) and obtain
\begin{align}
s^\pm_{f}(\omega)=& \int_{0}^{\frac{\pi}{2}} du \sin u   \int_{0}^{\frac{\pi}{2}}d\phi\; g^\pm_{f}(\omega \sin^{2} u \cos^{2} \phi)(kt^2g^\pm_{b}(\omega \sin^{2}u \sin^{2} \phi)g^{\mp}_b(\omega \cos^{2}u) \notag\\
&- J^2g^\pm_{f}(\omega \sin^{2}u \sin^{2} \phi)g^{\mp}_f(\omega \cos^{2}u) )\,,  \label{spmdefTJ-ferm}
\end{align}
and for bosons
\begin{align}
s^\pm_{b}(\omega)= - t^2\int_{0}^{\frac{\pi}{2}} du \sin u   \int_{0}^{\frac{\pi}{2}}d\phi\; g^\pm_{f}(\omega \sin^{2} u \cos^{2} \phi)g^\pm_{b}(\omega \sin^{2}u \sin^{2} \phi)g^{\mp}_f(\omega \cos^{2}u)\,,  \label{spmdefTJ-bos}
\end{align}
and we notice that $s_{\pm}(x)$ and $g_{\pm}(x)$ are  defined only for a positive argument.  Now it is left to find a real part $\Sigma'(\omega)$ of the self-energy. For this we use Kramers-Kronig transform
\begin{align}
\Sigma'(\omega) &= \dashint_{-\infty}^{+\infty} \frac{d\nu}{\pi} \frac{\Sigma''(\nu)-\Sigma''(\omega)}{\nu-\omega}\,.
\end{align}
Defining $\Sigma'^{(\pm)}_a(\omega)$ as $\Sigma_a'(\omega)=\Sigma'^{(+)}_a(\omega)\theta(\omega)+\Sigma'^{(-)}_a(-\omega)\theta(-\omega)$ we find
\begin{align}
\Sigma'^{(\pm)}_a(\omega) &= \pm \dashint_{0}^{+\infty} \frac{d\nu}{\pi}\left(\frac{\Sigma''^{(\pm)}_a(\nu)-\Sigma''^{(\pm)}_a(\omega)}{\nu-\omega}-\frac{\Sigma''^{(\mp)}_a(\nu)-\Sigma''^{(\pm)}_a(\omega)}{\nu+\omega}\right)\,.\label{SigmaprTJ}
\end{align}
At zero temperature we set chemical potential $\mu_a=\Sigma_a'(\omega =0)$, so introducing $h^{\pm}_a(\omega)$ as
\begin{align}
\Sigma_a'(\omega)-\Sigma_a'(0) = \begin{cases}
 4\sqrt{\omega} h^{+}_a(\omega), \quad \omega>0 \\
 4\sqrt{-\omega} h^{-}_a(-\omega), \quad \omega<0
\end{cases}\,. \label{defSigma1TJ}
\end{align}
and simplifying expressions we finally obtain
\begin{align}
h^{\pm}_a(\omega) = \pm \dashint_{0}^{+\infty} d\nu \left(\frac{\sqrt{\omega}s^{\pm}_a(\nu)-\sqrt{\nu}s^{\pm}_a(\omega)}{\sqrt{\nu}(\nu-\omega)}
+ \frac{\sqrt{\omega}s^{\mp}_a(\nu)+\sqrt{\nu}s^{\pm}_a(\omega)}{\sqrt{\nu}(\nu+\omega)}\right)\,. \label{hpmresTJ}
\end{align}
Now using the first pair of the Dyson-Schwinger equations \eqref{firstDS-ferm},\eqref{firstDS-bos} we can get $g^{\pm}_a$ from $s^{\pm}_a$ and $h^{\pm}_a$
\begin{align}
 g^{\pm}_a(\omega) = - \frac{ 4s^{\pm}_a(\omega)}{(4h^{\pm}_a(\omega)\mp \sqrt{\omega})^{2}+16 \pi^2(s^{\pm}_a(\omega))^{2}}\,. \label{gpmdef2TJ}
\end{align}
We solve Dyson-Schwinger equations iteratively using (\ref{spmdefTJ-ferm}), \eqref{spmdefTJ-bos}, (\ref{hpmresTJ}) and (\ref{gpmdef2TJ}) and also imposing the initial conditions
\begin{align}
&g^{\pm}_a(0) = \frac{C_{a}}{\pi J} \sin(\frac{\pi}{4}\pm \theta_{a}), \quad 
s_{a}^{\pm}(0) = -\frac{1}{4\pi C_{a}} \sin(\frac{\pi}{4}\pm \theta_{a}), \quad h_{a}^{\pm}(0) = \mp \frac{1}{4C_{a}}\cos(\frac{\pi}{4}\pm \theta_{a})\,.
\end{align}
where the constants $C_{a}$ are given in (\ref{eq:CfCbsol}).

\subsection{Luttinger constraints and fermionic charge vs chemical potential}
\begin{figure}[h!]
\includegraphics[width=0.47\textwidth]{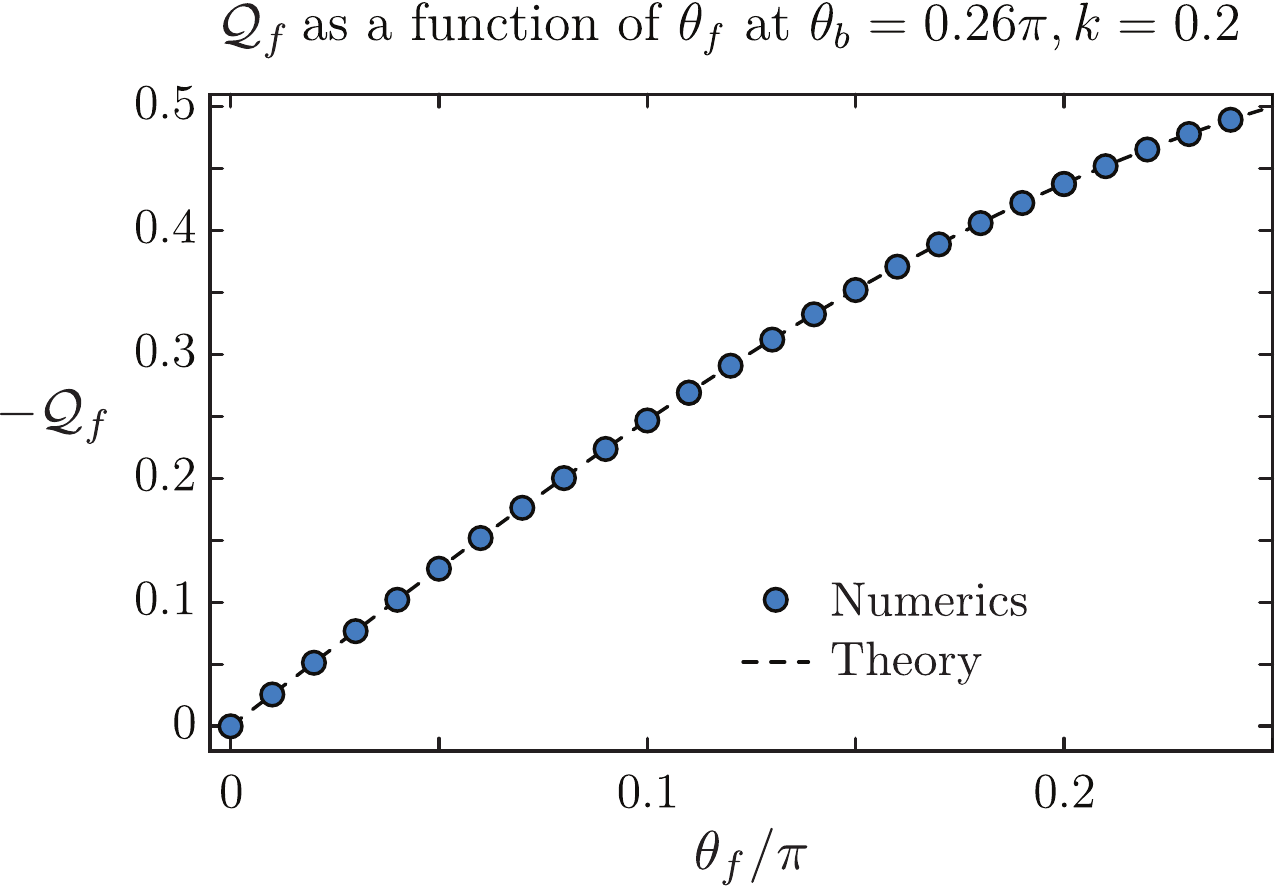}\quad\quad
\includegraphics[width=0.47\textwidth]{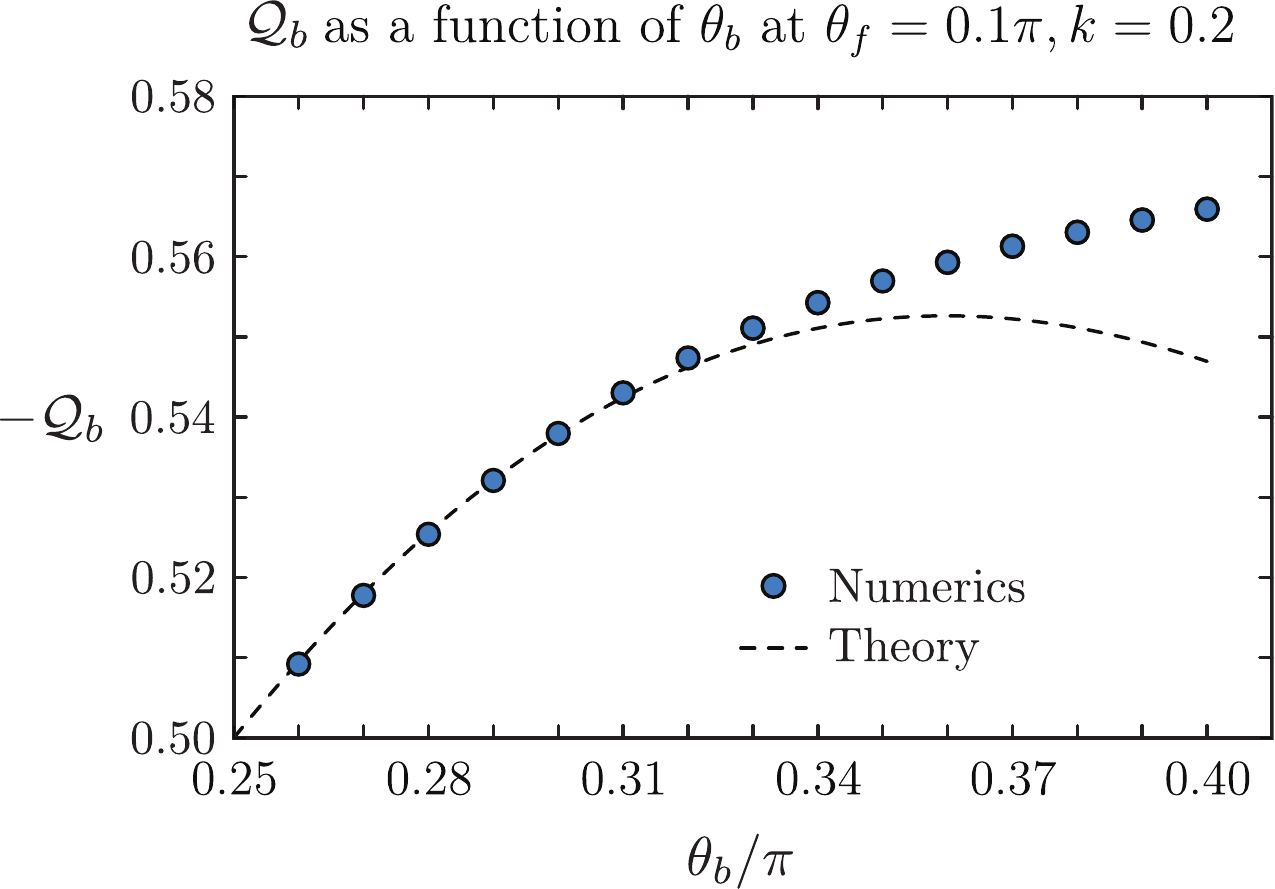}
\caption{The Luttinger constraints for fermions (left) and bosons (right).  Blue dots are numerically computed Luttinger constraints and the dashed black lines are the analytical functions \eqref{eq:Q},\eqref{eq:S}. Here we fix the parameters $k=0.2$, and $\theta_b=0.26\pi$ (left), $\theta_f=0.1\pi$ (right). We note that the precision for each point is in the range of $10^{-5}-10^{-7}$.\label{Fig:LuttingConstr}}
\end{figure}

We can numerically investigate the solution similarly to the previous paper \cite{Tikhanovskaya:2020elb} computing the Luttinger constraints \eqref{eq:Ltfdel14},\eqref{eq:Ltbdel14} both numerically and analytically that we rewrite as follows
\begin{eqnarray}
 \mathcal{Q}_{f}(\theta_f)&=&-\frac{\theta_f}{\pi}-\frac14{\sin(2\theta_f)} = \left\langle f^\dagger f \right\rangle -\frac{1}{2} \,, \label{eq:Q}\\
 \mathcal{Q}_{b}(\theta_b)&=& -\frac{\theta_b}{\pi}-\frac14{\sin(2\theta_b)} = -\frac{1}{2}- \left\langle b^\dagger b \right\rangle\,. \label{eq:S}
\end{eqnarray}
Numerically we compute the areas under fermionic and bosonic spectral density curves respectively at fixed parameters. We show the comparison of the analytical and numerical results in the Fig.\ref{Fig:LuttingConstr}.

\begin{figure}[h!]
\includegraphics[width=0.45\textwidth]{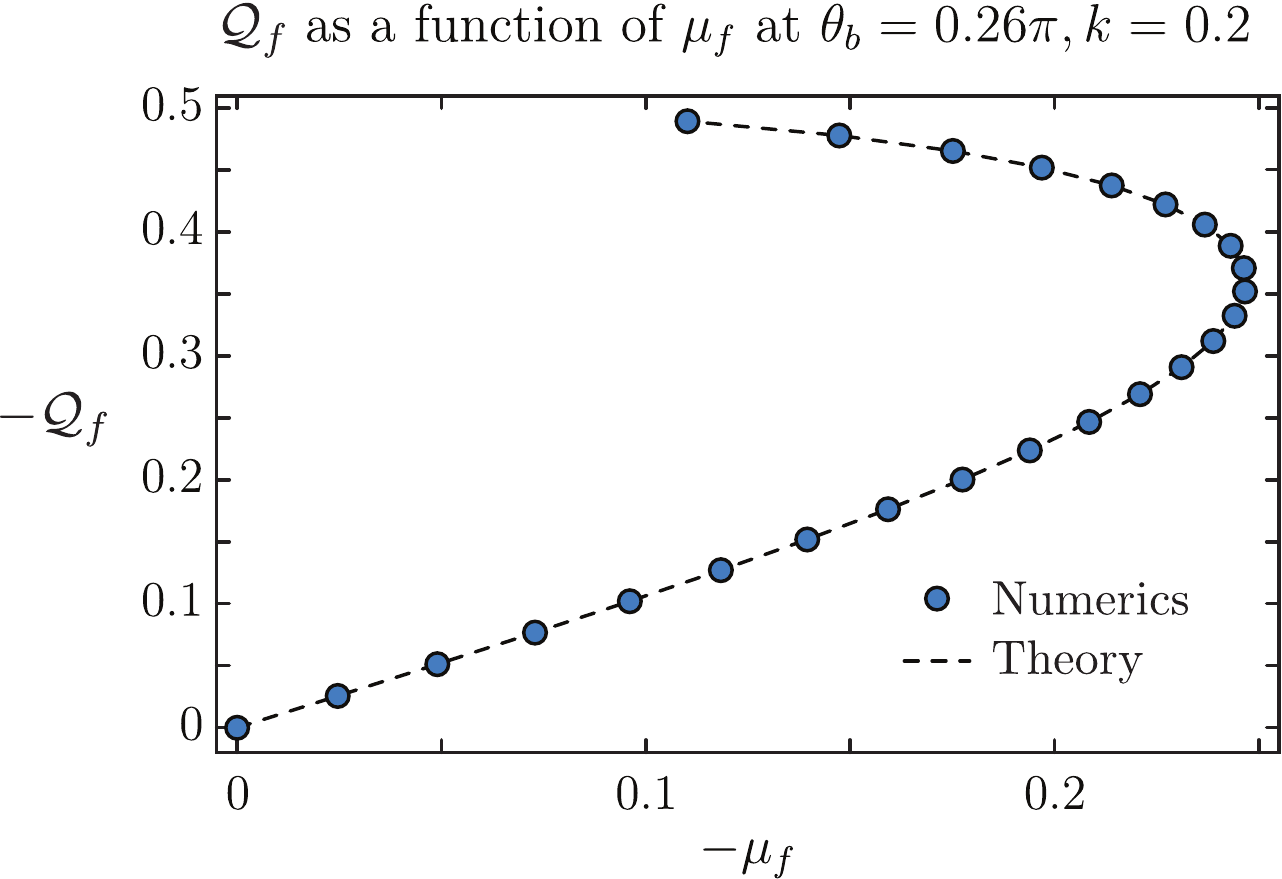}\quad\quad
\includegraphics[width=0.46\textwidth]{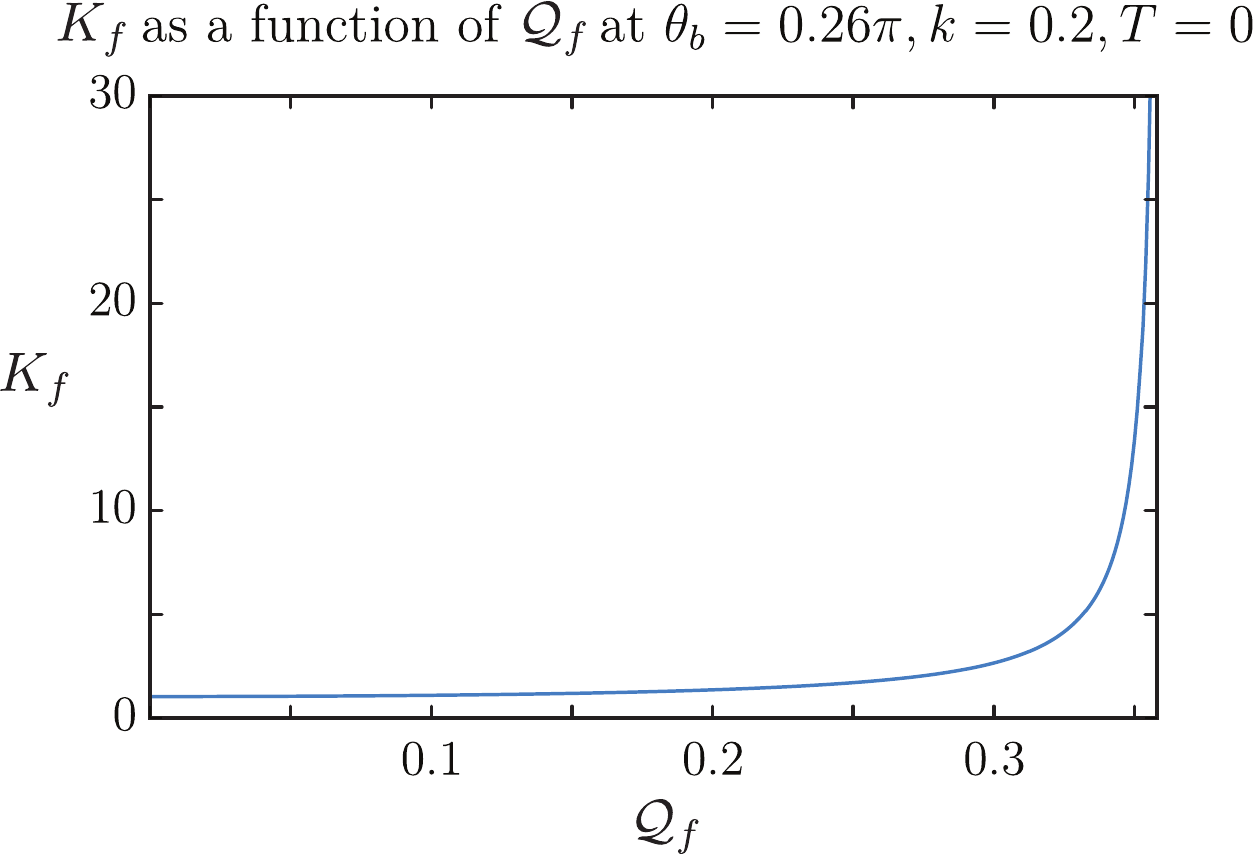}
\caption{Left: Charge as a function of fermionic chemical potential at $T=0$. The function $\mathcal{Q}_{f}$ depends on $\mu_f$ through $\theta_f$. Right: Compressibility as a function of charge at $T=0$. Parameters for both plots:  $\theta_b=0.26\pi$, $k=0.2$, $J=1$ and $t=1$. Dashed line on the left plot is numerical line connecting the blue points. The divergence of compressibility happens at $\theta_f\sim 0.15\pi$.\label{Qvsmu}}
\end{figure}

In addition to the Luttinger constraints, we numerically compute the fermionic charge as a function of the chemical potential that is given by the real part of the self energy at zero frequency $\mu=\Sigma'(\omega=0)$. The result is presented in the Fig.\ref{Qvsmu}.

The behavior of the fermionic charge as a function of chemical potential is similar to the SYK case \cite{Tikhanovskaya:2020elb}. We do not see special features in the bosonic charge $\mathcal{Q}_{b}$.

\section{Corrections to Conformality}\label{app:kernel}

The notation we adopt here is the same as \cite{Gu:2019jub, Guo:2020aog, Tikhanovskaya:2020elb}, where we represent the functions as a four-component vector indexed by boson/fermion and $\pm\tau$ branches. We use the following basis for the UV-source and the corrections of Green's function and self-energy:
 \begin{equation}\label{eq:sigmaansatz}
   \sigma_h(\tau)=\begin{pmatrix}
                   \delta \sigma_{b+} \Sigma_{b}^c(\tau) \\
                   \delta \sigma_{b-} \Sigma_{b}^c(\tau)\\
                   \delta \sigma_{f+} \Sigma_{f}^c(\tau)\\
                   \delta \sigma_{f-} \Sigma_{f}^c(\tau)
                 \end{pmatrix}|J\tau|^{1-h},
 \end{equation}
\begin{equation}\label{eq:ansatz}
  \delta G(\tau)=\begin{pmatrix}
                   \delta G_{b+} G_{b}^c(\tau) \\
                   \delta G_{b-} G_{b}^c(\tau)\\
                   \delta G_{f+} G_{f}^c(\tau)\\
                   \delta G_{f-} G_{f}^c(\tau)
                 \end{pmatrix}|J\tau|^{1-h},
                 \qquad
    \delta \Sigma(\tau)=\begin{pmatrix}
                   \delta \Sigma_{b+} \Sigma_{b}^c(\tau) \\
                   \delta \Sigma_{b-} \Sigma_{b}^c(\tau)\\
                   \delta \Sigma_{f+} \Sigma_{f}^c(\tau)\\
                   \delta \Sigma_{f-} \Sigma_{f}^c(\tau)
                 \end{pmatrix}|J\tau|^{1-h}.
\end{equation}
Here $\pm$ means positive/negative $\tau$ branches, and $b,f$ denote boson and fermion respectively.

\subsubsection{Kernel of the $t$-$J$ Model in the bosonic spinon $+$ fermionic holon case }
  In this part we compute the kernel $K_G$ of the $t$-$J$ model in bosonic spinon $+$ fermionic holon case. The basic ingredient is $W_\Sigma$ and $W_G$ as defined in \cite{Gu:2019jub,Guo:2020aog}. We elaborate the definition by explicitly writing down on indices
\begin{equation}
    \begin{aligned}
    W_{\Sigma} (\tau_1,\tau_2;\tau_3,\tau_4)_{a\tilde{a}} &=\left. \frac{\delta G_{*a}(\tau_1,\tau_2)}{\delta \Sigma_{\tilde{a}}(\tau_3,\tau_4)}\right\vert_{\Sigma^c}, \\
    W_{G} (\tau_1,\tau_2;\tau_3,\tau_4)_{a\tilde{a}} &= \left.\frac{\delta \Sigma_{*a}(\tau_1,\tau_2)}{\delta G_{\tilde{a}}(\tau_3,\tau_4)}\right\vert_{G^c},
    \end{aligned}
    \label{Eq:Windex}
\end{equation}where $a,\tilde{a}=\ff,\fb$ denote fermion and boson. Using Eq.\eqref{Eq:Windex} and the conformal saddle point equations, we obtain
\begin{equation}\label{}
  W_\Sigma(1,2;3,4)=\begin{pmatrix}
                      G_\ff^{13}G_\ff^{42} & 0 \\
                      0 & G_\fb^{13}G_\fb^{42}
                    \end{pmatrix},
\end{equation} and diagramatically
\begin{equation}\label{eq:Wsigmagraph2}
  W_\Sigma(1,2;3,4)=\begin{pmatrix}
                      \begin{tikzpicture}[baseline={([yshift=-4pt]current bounding box.center)}]
                     \draw[thick, dashed, mid arrow] (40pt,12pt)--(0pt,12pt);
                     \draw[thick, dashed, mid arrow] (0pt,-12pt)--(40pt,-12pt);
                     \node at (-5pt,12pt) {\scriptsize $1$};
                     \node at (-5pt,-12pt) {\scriptsize $2$};
                     \node at (48pt,12pt) {\scriptsize $3$};
                     \node at (48pt,-12pt) {\scriptsize $4$};
                     \end{tikzpicture}
                        & 0 \\
                      0 &
                      \begin{tikzpicture}[baseline={([yshift=-4pt]current bounding box.center)}]
                     \draw[thick, mid arrow] (40pt,12pt)--(0pt,12pt);
                     \draw[thick, mid arrow] (0pt,-12pt)--(40pt,-12pt);
                     \node at (-5pt,12pt) {\scriptsize $1$};
                     \node at (-5pt,-12pt) {\scriptsize $2$};
                     \node at (48pt,12pt) {\scriptsize $3$};
                     \node at (48pt,-12pt) {\scriptsize $4$};
                     \end{tikzpicture}
                    \end{pmatrix}.
\end{equation}
Here we use dashed line to denote fermion and solid line to denote boson.

$W_G$ differs from the other scheme by a switching $\fb\leftrightarrow f,~\ff\leftrightarrow b$ and an overall minus sign.
\begin{equation}\label{}
 W_G(1,2;3,4)=
               \begin{pmatrix}
                 t^2 G_\fb^{12}G_\fb^{21} \delta^{24;13}& t^2 G_\fb^{43} G_\ff^{12}(\delta^{24;13}+\delta^{23;14})   \\
                 -kt^2 G_\ff^{43}G_\fb^{12}(\delta^{24;13}+\delta^{23;14})
                 &
                 -kt^2 G_\ff^{12} G_\ff^{21}\delta^{24;13}+J^2\left(2 G_\fb^{12}G_\fb^{21}\delta^{24;13}+(G_\fb^{12})^2\delta^{23;14}\right)
               \end{pmatrix}
\end{equation}
where notation $\delta^{ij;kl}=\delta(\tau_i-\tau_j)\delta(\tau_k-\tau_l)$. Diagrammatically, $W_G$ has the following representation
\begin{equation}\label{}
  W_G(1,2;3,4)=\begin{pmatrix}
        t^2
        \begin{tikzpicture}[baseline={([yshift=-4pt]current bounding box.center)}]
        \draw[thick, densely dotted] (10pt,15pt)--(0pt,15pt);
        \draw[thick, densely dotted] (0pt,-15pt)--(10pt,-15pt);
        \draw[thick, mid arrow] (0pt,15pt)..controls (5pt,7pt) and (5pt,-7pt)..(0pt,-15pt);
        \draw[thick, mid arrow] (0pt,-15pt)..controls (-5pt,-7pt) and (-5pt,7pt)..(0pt,15pt);
        \node at (-5pt,15pt) {\scriptsize $1$};
        \node at (-5pt,-15pt) {\scriptsize $2$};
        \node at (18pt,15pt) {\scriptsize $3$};
        \node at (18pt,-15pt) {\scriptsize $4$};
        \end{tikzpicture}
& t^2\left(
         \begin{tikzpicture}[baseline={([yshift=-4pt]current bounding box.center)}]
        \draw[thick, densely dotted] (10pt,15pt)--(0pt,15pt);
        \draw[thick, densely dotted] (0pt,-15pt)--(10pt,-15pt);
        \draw[thick,  mid arrow] (0pt,15pt)..controls (5pt,7pt) and (5pt,-7pt)..(0pt,-15pt);
        \draw[thick, dashed, mid arrow] (0pt,-15pt)..controls (-5pt,-7pt) and (-5pt,7pt)..(0pt,15pt);
        \node at (-5pt,15pt) {\scriptsize $1$};
        \node at (-5pt,-15pt) {\scriptsize $2$};
        \node at (18pt,15pt) {\scriptsize $3$};
        \node at (18pt,-15pt) {\scriptsize $4$};
        \end{tikzpicture}
        +
        \begin{tikzpicture}[baseline={([yshift=-4pt]current bounding box.center)}]
        \draw[thick, densely dotted] (0pt,15pt)--(20pt,-15pt);
        \draw[thick, densely dotted] (20pt,15pt)--(0pt,-15pt);
        \draw[thick,  mid arrow] (0pt,-15pt)..controls (5pt,-7pt) and (5pt,7pt)..(0pt,15pt);
        \draw[thick, dashed,mid arrow] (0pt,-15pt)..controls (-5pt,-7pt) and (-5pt,7pt)..(0pt,15pt);
        \node at (-5pt,15pt) {\scriptsize $1$};
        \node at (-5pt,-15pt) {\scriptsize $2$};
        \node at (28pt,15pt) {\scriptsize $3$};
        \node at (28pt,-15pt) {\scriptsize $4$};
        \end{tikzpicture}\right)  \vspace{1em}\\
        -kt^2\left(
         \begin{tikzpicture}[baseline={([yshift=-4pt]current bounding box.center)}]
        \draw[thick, densely dotted] (10pt,15pt)--(0pt,15pt);
        \draw[thick, densely dotted] (0pt,-15pt)--(10pt,-15pt);
        \draw[thick, dashed, mid arrow] (0pt,15pt)..controls (5pt,7pt) and (5pt,-7pt)..(0pt,-15pt);
        \draw[thick, mid arrow] (0pt,-15pt)..controls (-5pt,-7pt) and (-5pt,7pt)..(0pt,15pt);
        \node at (-5pt,15pt) {\scriptsize $1$};
        \node at (-5pt,-15pt) {\scriptsize $2$};
        \node at (18pt,15pt) {\scriptsize $3$};
        \node at (18pt,-15pt) {\scriptsize $4$};
        \end{tikzpicture}
        +
        \begin{tikzpicture}[baseline={([yshift=-4pt]current bounding box.center)}]
        \draw[thick, densely dotted] (0pt,15pt)--(20pt,-15pt);
        \draw[thick, densely dotted] (20pt,15pt)--(0pt,-15pt);
        \draw[thick, dashed, mid arrow] (0pt,-15pt)..controls (5pt,-7pt) and (5pt,7pt)..(0pt,15pt);
        \draw[thick, mid arrow] (0pt,-15pt)..controls (-5pt,-7pt) and (-5pt,7pt)..(0pt,15pt);
        \node at (-5pt,15pt) {\scriptsize $1$};
        \node at (-5pt,-15pt) {\scriptsize $2$};
        \node at (28pt,15pt) {\scriptsize $3$};
        \node at (28pt,-15pt) {\scriptsize $4$};
        \end{tikzpicture}\right)
        \hspace{10pt}
    & -kt^2
        \begin{tikzpicture}[baseline={([yshift=-4pt]current bounding box.center)}]
        \draw[thick, densely dotted] (10pt,15pt)--(0pt,15pt);
        \draw[thick, densely dotted] (0pt,-15pt)--(10pt,-15pt);
        \draw[thick, dashed,mid arrow] (0pt,15pt)..controls (5pt,7pt) and (5pt,-7pt)..(0pt,-15pt);
        \draw[thick, dashed,mid arrow] (0pt,-15pt)..controls (-5pt,-7pt) and (-5pt,7pt)..(0pt,15pt);
        \node at (-5pt,15pt) {\scriptsize $1$};
        \node at (-5pt,-15pt) {\scriptsize $2$};
        \node at (18pt,15pt) {\scriptsize $3$};
        \node at (18pt,-15pt) {\scriptsize $4$};
        \end{tikzpicture}
        +J^2
        \left(2
         \begin{tikzpicture}[baseline={([yshift=-4pt]current bounding box.center)}]
        \draw[thick, densely dotted] (10pt,15pt)--(0pt,15pt);
        \draw[thick, densely dotted] (0pt,-15pt)--(10pt,-15pt);
        \draw[thick,  mid arrow] (0pt,15pt)..controls (5pt,7pt) and (5pt,-7pt)..(0pt,-15pt);
        \draw[thick, mid arrow] (0pt,-15pt)..controls (-5pt,-7pt) and (-5pt,7pt)..(0pt,15pt);
        \node at (-5pt,15pt) {\scriptsize $1$};
        \node at (-5pt,-15pt) {\scriptsize $2$};
        \node at (18pt,15pt) {\scriptsize $3$};
        \node at (18pt,-15pt) {\scriptsize $4$};
        \end{tikzpicture}
        +
        \begin{tikzpicture}[baseline={([yshift=-4pt]current bounding box.center)}]
        \draw[thick, densely dotted] (0pt,15pt)--(20pt,-15pt);
        \draw[thick, densely dotted] (20pt,15pt)--(0pt,-15pt);
        \draw[thick,  mid arrow] (0pt,-15pt)..controls (5pt,-7pt) and (5pt,7pt)..(0pt,15pt);
        \draw[thick, mid arrow] (0pt,-15pt)..controls (-5pt,-7pt) and (-5pt,7pt)..(0pt,15pt);
        \node at (-5pt,15pt) {\scriptsize $1$};
        \node at (-5pt,-15pt) {\scriptsize $2$};
        \node at (28pt,15pt) {\scriptsize $3$};
        \node at (28pt,-15pt) {\scriptsize $4$};
        \end{tikzpicture}\right)  \vspace{1em}
      \end{pmatrix},\label{eq:WGgraphic2}
\end{equation}where a dotted line without arrow represents a $\delta$-function. In the basis given by \eqref{eq:sigmaansatz} and \eqref{eq:ansatz} $W_\Sigma$ and $W_\Sigma$ take the familiar form
\begin{equation}
\begin{split}
 W_\Sigma(h)&=w(\Delta_\ff,\theta_\ff;h)\oplus w(\Delta_\fb,\theta_\fb;h),\\
w(\Delta,\theta;h)&=\frac{\Gamma(2\Delta-1+h)\Gamma(2\Delta-h)}{\Gamma(2\Delta)\Gamma(2\Delta-1)\sin(2\pi\Delta)}
    \begin{pmatrix}
     \sin(\pi h+2\theta) & -\sin(2\pi\Delta)+\sin(2\theta) \\
    -\sin(2\pi\Delta)-\sin(2\theta) & \sin(\pi h-2\theta)
    \end{pmatrix},
\end{split}
\end{equation}
and
\begin{equation}\label{}
  W_G(h)=\left(
\begin{array}{cccc}
 1 & 0 & 1 & 1 \\
 0 & 1 & 1 & 1 \\
 -k\frac{\cos 2\theta_\ff}{\cos 2 \theta_\fb} & -k\frac{\cos 2\theta_\ff}{\cos 2 \theta_\fb}  & k \frac{\cos 2\theta_\ff}{\cos 2 \theta_\fb}+2 & k \frac{\cos 2\theta_\ff}{\cos 2 \theta_\fb}+1 \\
 -k\frac{\cos 2\theta_\ff}{\cos 2 \theta_\fb} & -k\frac{\cos 2\theta_\ff}{\cos 2 \theta_\fb} & k \frac{\cos 2\theta_\ff}{\cos 2 \theta_\fb}+1 & k \frac{\cos 2\theta_\ff}{\cos 2 \theta_\fb}+2 \\
\end{array}
\right).
\end{equation}
Finally $K_G(h)=W_\Sigma(h)W_G(h)$ and $K_\Sigma(h)=W_G(h)W_\Sigma(h)$.
It turns out that in this scheme the kernel is identical to that of the fermionic spinon + bosonic holon theory if we substitute $\theta_f\to \theta_\fb$, $\theta_b\to \theta_\ff$.

\subsubsection{Linear order correction at zero temperature}
  Following derivations in \cite{Tikhanovskaya:2020elb}, the first order conformal corrections at zero temperature take the following form :
\begin{equation}\label{eq:deltaGalpha}
\begin{split}
     \delta G(\tau)&=\sum_h \delta_h G(\tau)\,,\\
     \delta_h G(\tau)&=-\alpha_{h}v_{h}\frac{G^c(\tau)}{|J\tau|^{h-1}}\,,\quad \alpha_{h}=-\frac{w_{h} W_\Sigma(h)\vec{\sigma}_{h}}{ k_G'(h)}\,.
\end{split}
\end{equation} Here the sum is over the operator spectrum of the $t$-$J$ model determined by \eqref{eq:det(1-K)=0}, and $v_{h}$ and $w_{h}$ are the corresponding right and left eigenvectors of $K_{G}(h)$ respectively which have eigenvalue $k_G(h)=1$ and normalized as $w_{h}v_{h}=1$, and $\vec{\sigma}_h=(\delta\sigma_{b+},\delta\sigma_{b-},\delta\sigma_{f+},\delta\sigma_{f-})^T$.

It can be verified that the eigenspace corresponding to the two $h=1$ operators are spanned by $\delta G^c_a/\delta\theta_b$ and $\delta G^c_a/\delta\theta_b$ (i.e. changes of $\theta_b$ and $\theta_f$) which by the Luttinger relations \eqref{eq:Ltf},\eqref{eq:LtB} are related to $\delta\mathcal{Q}_f$ and $\delta\mathcal{Q}_b$ respectively. Therefore, there is no need to include these corrections if we use the renormalized value of $\theta_f,\theta_b$.

\subsubsection{Nonlinear order correction at zero temperature}

  It is also possible to obtain corrections of higher order in $\alpha$ following derivations in \cite{Tikhanovskaya:2020elb}. The result takes the form
\begin{align}
\label{GallcorrAppendix}
G(\tau) &=G^{c}(\tau)  \bigg(1-\sum_{h}^{}\frac{\alpha_{h} v_{h}}{|J\tau|^{h-1}}-\sum_{h,h'}^{}\frac{a_{hh'}\alpha_{h}\alpha_{h'}}{|J\tau|^{h+h'-2}}-\sum_{h,h',h''}^{}\frac{a_{hh'h''}\alpha_{h}\alpha_{h'}\alpha_{h''}}{|J\tau|^{h+h'+h''-3}}-\dots\bigg)\,,
\end{align}
where  $v_{h}$, $a_{hh'}$, $a_{hh'h''}$, etc are four-component vectors. The coefficients $a_{hh'}$, $a_{hh'h''}$ can be calculated using the recursion procedure described in \cite{Tikhanovskaya:2020elb}. As an example, we perform a calculation for $a_{hh'}$, using Eq. (4.32) of \cite{Tikhanovskaya:2020elb}:
\begin{equation}\label{}
  \delta^2 G = \frac{1}{1-W_\Sigma W_G}\left[W_\Sigma \bar{\delta}^2 \Sigma_*[G]+\bar{\delta}^2 G_*[\Sigma]\right], \label{eq:deltakG}
\end{equation}where $\Sigma_*[G]$ and $G_*[\Sigma]$ denote the RHS of Schwinger-Dyson equations \eqref{Eq:EoM1}-\eqref{Eq:EoM4} with $(i\omega_n+\mu)$ dropped. The meaning of $\bar{\delta}^2\Sigma_*[G]$ is that we compute the second-order variation of $\Sigma_*$ using only first-order variation of $G$.

The second term in the bracket of \eqref{eq:deltakG} can be computed using the same procedure as \cite{Tikhanovskaya:2020elb}, with the result
\begin{align}\label{eq:d2G}
\frac{\bar{\delta}^{2}G_{*}(\tau)}{G^{c}(\tau)} =  \sum_{h,h'} F(h+h'-1)^{-1}\big(F(h)v_{h}\cdot F(h')v_{h'}\big)\frac{\alpha_{h}\alpha_{h'}}{|J\tau|^{h+h'-2}}\,,
\end{align}
where ``$\cdot$" means multiply two vectors entry-wise and $F(h)$ is a 4 by 4 matrix that acts on $v_h$:
\begin{equation}\label{}
\begin{split}
 F(h)&= f(h,\theta_b)\oplus f(h,\theta_f),\\
 f(h,\theta) &= -\frac{\Gamma(3/2-h)}{2\sqrt{\pi}}\left(
\begin{array}{cc}
 -e^{-\frac{1}{2} i \pi  h} \left(e^{2 i \theta }+i\right) & -e^{\frac{i \pi  h}{2}} \left(e^{2 i \theta }-i\right) \\
 i \left(e^{2 i \theta }+i\right) e^{\frac{1}{2} i (\pi  h-4 \theta )} & \left(-1-i e^{2 i \theta }\right)
   e^{-\frac{1}{2} i (\pi  h+4 \theta )} \\
\end{array}
\right).
\end{split}
\end{equation}

The first term in the bracket of \eqref{eq:deltakG} can be computed by expanding the Schwinger-Dyson equation $\Sigma=\Sigma_*[G]$ to second order, with the result

\begin{equation}\label{eq:d2sigma}
  \frac{\bar{\delta}^2 \Sigma_*(\tau)}{\Sigma^c(\tau)}=\sum_{h,h'}\frac{\alpha_h\alpha_{h'}}{|J\tau|^{h+h'-2}}d_{hh'}\,,
\end{equation} and $d_{hh'}$ is a four-component vector whose entries are
\begin{eqnarray}
  d_{hh';b+} &=& \frac{1}{2} (v_{hb+} (v_{h'f-}+v_{h'f+})+v_{hf-} (v_{h'b+}+v_{h'f+})+v_{hf+} (v_{h'b+}+v_{h'f-}))\,, \\
  d_{hh';b-} &=& \frac{1}{2} (v_{hb-} (v_{h'f-}+v_{h'f+})+v_{hf-} (v_{h'b-}+v_{h'f+})+v_{hf+} (v_{h'b-}+v_{h'f-}))\,,  \\
  d_{hh';f+} &=&  \frac{1}{2} \eta  \Big(v_{hb-} (v_{h'b+}+v_{h'f+})+v_{hb+} (v_{h'b-}+v_{h'f+})-2 v_{hf-} v_{h'f+}\\\nonumber
  &&+v_{hf+} (v_{h'b-}+v_{h'b+}-2 (v_{h'f-}+v_{h'f+}))\Big)+v_{hf-} v_{h'f+}+v_{hf+} (v_{h'f-}+v_{h'f+})  \\
  d_{hh';f-} &=&  \frac{1}{2} \eta  \Big(v_{hb-} (v_{h'b+}+v_{h'f-})+v_{hb+} (v_{h'b-}+v_{h'f-})-2 v_{hf+} v_{h'f-}\,,\\
  &&+v_{hf-} (v_{h'b-}+v_{h'b+}-2 (v_{h'f-}+v_{h'f+}))\Big)+v_{hf-} (v_{h'f-}+v_{h'f+})+v_{hf+} v_{h'f-}\,. \nonumber
\end{eqnarray}
We remind the reader that $\eta= -k \cos(2\theta_b)/\cos(2\theta_f)$. Combining \eqref{eq:d2G} and \eqref{eq:d2sigma}, we obtain
\begin{align}
a_{hh'}=& -\big(1-W_{\Sigma}(h+h'-1)W_{G}\big)^{-1} \Big( F(h+h'-1)^{-1}\big(F(h)v_{h}\cdot F(h')v_{h'}\big)+W_\Sigma(h+h'-1)d_{hh'}\Big)\,. \label{ah1h2}
\end{align}

\section{Qualitative analysis of potentially relevant operators in the $t$-$J$ model}\label{app:diagram}

The goal of this analysis is to determine the range of parameters regarding the existence and stability of the saddle conformal solution, and to study the relevance of time-reparameterization mode $h=2$. The outcome of the analysis is in Fig.~\ref{fig:diagram_fermionic_spin} and Fig.~\ref{fig:bosonic_spin_plot}. We will mainly discuss the fermionic spinon + bosonic holon model. The analysis of the fermionic holon + bosonic spinon model is completely parallel as the kernel is related to by $\theta_f\to \theta_\fb$, $\theta_b\to \theta_\ff$ .

We will be analyzing the function
\begin{equation}\label{eq:det1-K_G=0}
  F(h)\equiv\det(K_G(h)-1)=0,
\end{equation}whose zeroes determine the scaling dimension of irrelevant operators appearing in the corrections.
The conformal saddle exists if the Eq.\eqref{eq:CfCb} has real solutions for $C_f$ and $C_b$, which implies
\begin{equation}\label{eq:existence1}
  \cos(2\theta_f)+k\cos(2\theta_b)>0.
\end{equation} Here $\theta_f\in[-\pi/4,\pi/4]$ and $\theta_b\in[\pi/4,\pi/4]$ are assumed to respect unitarity constraint.

By numerically inspecting the function $F(h)$, we found that there is almost always a solution with scaling dimension $\Re~h_1<\frac{3}{2}$. Depending on parameters, this solution may be on the real axis within the range $\frac{1}{2}<h_1<\frac{3}{2}$, or it may move onto the imaginary axis such that $h_1=1/2\pm is,~s>0$.

Qualitative behaviors of the solution can be described by simple inequalities, which we describe below. First, whether the solution $h_1$ is real or complex depends on the sign of $F(h)$ near the pole $h=1/2$. When $h_1$ moves from real to complex, the pole switches sign. Therefore the existence of a complex solution corresponds to
\begin{equation}\label{eq:bound1}
\begin{split}
   0> & 16\pi^2\lim_{h\to 1/2}(h-\frac{1}{2})^4 F(h)=8 \left(4 \cos \left(2 \theta _b\right)+\pi \right) \cos \left(2 \theta _f\right)+16 k \cos \left(4 \theta _b\right)+16 k+\pi  (3 \pi -4)\\
     & +2 (\pi -2) k \cos \left(2 \theta _b\right) \left(8 \cos \left(2 \theta _b\right)+3 \pi -4\right) \sec \left(2 \theta _f\right)+4 (3 \pi -4) (k+1) \cos \left(2 \theta _b\right).
\end{split}
\end{equation}
Second, generically $F(h)$ has a double zero at $h=1$, corresponding two gauge charges $\mathcal{Q}_f,~\mathcal{Q}_b$ which have scaling dimension 1. When $h_*$ crosses 1, $F(1)$ will be a triple zero such that
\begin{equation}\label{}
  0=F''(1)=16 \left(\pi  \cos \left(2 \theta _b\right)+2\right) \left(\pi  \cos \left(2 \theta _f\right)+2\right) \sec \left(2 \theta _f\right) \left(k \cos \left(2 \theta _b\right)+\cos \left(2 \theta _f\right)\right).
\end{equation}  Assuming, \eqref{eq:existence1} holds, the only factor that could change sign is $2+\pi\cos(2\theta_b)$, so the condition for $1/2<h_1<1$ is
\begin{equation}\label{eq:bound2}
  \theta_b>\frac{1}{2}\arccos\left(\frac{-2}{\pi}\right)\approx 0.3598\pi.
\end{equation}We also remark that this is the maximum of the LHS of Luttinger constraint \eqref{eq:LtB}. When $h_1=\frac{3}{2}$, it will be absorbed by the pole of $F(h)$ at $h=\frac{3}{2}$, so the mode disappears (it has zero OPE with the $f$ and $b$). Physically, this corresponds to $\theta_b=\pi/4$, i.e. the holon density is zero.

Additionally, there can be a second solution near $h_{2}$ near $h=2$. When this mode crosses $h=2$, $F'(2)$ changes sign. The condition that $h_{2}<2$ is
\begin{equation}\label{eq:bound3}
  0>F'(2)=\frac{16}{81} \sec \left(2 \theta _f\right) \left(-4 k \cos \left(2 \theta _b\right)-3 \pi  k \cos \left(4 \theta _b\right)+4 \cos \left(2 \theta _f\right)+3 \pi  \cos \left(4 \theta _f\right)-3 \pi  k+3 \pi \right).
\end{equation}
The above constraints define all the boundaries in Fig.~\ref{fig:diagram_fermionic_spin}. Same inequalties with $\theta_f\to \theta_\fb$, $\theta_b\to \theta_\ff$  define the boundaries in Fig.~\ref{fig:bosonic_spin_plot}.

\section{Calculation of Optical Conductivity}\label{sec:sigmah=2}

In this appendix we give details on the evaluation of \eqref{eq:sigmah=2}. For convenience, we will work in units where $\beta=2\pi$.

Our starting point is the Kubo formula \eqref{sigmaomega}, recast into the following form
\begin{equation}\label{}
  \Re\sigma(\omega)=\frac{\sigma_0 }{ C_e^2/J^2}\int \rd \Omega\frac{1}{\omega} \left(\frac{1}{e^{2\pi \Omega}+1}-\frac{1}{e^{2\pi (\Omega+\omega)}+1}\right)\rho_e(\Omega)\rho_e(\Omega+\omega)\,,
\end{equation} 
Here the prefactors are chosen such that the conformal contribution is $\sigma_0$.

The electron spectral weight is given in \eqref{eq:rhoefinal}, 
\begin{equation}\label{}
  \rho_e(\omega)=\rho_e^c(\omega)\left(1-\sum_{h}V_h^\mathrm{A} \mathcal{R}_{h}^{\textrm{A}}\big(\omega-\calE_{e}\big)-V_h^\mathrm{S}\mathcal{R}_{h}^{\textrm{S}}\big( \omega-\calE_{e}\big)\right),
\end{equation} where 
\begin{equation}\label{}
  V_h^\mathrm{A/S}=\frac{\alpha_{h}}{2 (\beta J)^{h-1}}(v_{hf+}\pm v_{hf-}\pm v_{hb+}+v_{hb-})\,.
\end{equation}

The conformal contribution is
\begin{equation}\label{}
  \Re \sigma_0(\omega)=\sigma_0 I(0^+,\omega,\ce_e)\,,
\end{equation} 
where the integral involved is of the type
\begin{equation}\label{}
  I(\epsilon,\omega,\ce)=\int \rd \Omega\frac{e^{\epsilon\Omega}}{ \omega} \left(\frac{1}{e^{2\pi \Omega}+1}-\frac{1}{e^{2\pi (\Omega+\omega)}+1}\right)\frac{\cosh(\pi\Omega)\cosh(\pi(\Omega+\omega))}{\cosh(\pi(\Omega-\ce))\cosh(\pi(\Omega+\omega-\ce))}.
\end{equation}
We have add a convergence factor $e^{\epsilon \Omega}$ with $\epsilon\to 0^+$ eventually, and then split the two terms in the parenthesis. The individual integrands are periodic under $\Omega\to\Omega+i$, so we can use an rectangular contour between $\Im \Omega=0$ and $\Im \Omega=1$. The encircled poles are at $\Omega=\ce+\frac{i}{2}$ and $\Omega=\omega+\ce+\frac{i}{2}$. We obtain
\begin{equation}\label{}
  I(\epsilon,\omega,\ce)=\frac{1-e^{-\epsilon\omega}}{2\omega\sin\frac{\epsilon}{2}}e^{\epsilon\ce},
\end{equation}
\begin{equation}\label{}
  \Re \sigma_0(\omega)=\sigma_0I(0^+,\omega,\ce)=\sigma_0,
\end{equation} which is independent of frequency as expected from scaling analysis.

Next we look at contribution due to corrections.
\begin{equation}\label{}
\begin{split}
  \Re \delta_h \sigma(\omega)&=-\sigma_0\int \rd \Omega\frac{e^{\epsilon\Omega}}{ \omega} \left(\frac{1}{e^{2\pi \Omega}+1}-\frac{1}{e^{2\pi (\Omega+\omega)}+1}\right)\frac{\cosh(\pi\Omega)\cosh(\pi(\Omega+\omega))}{\cosh(\pi(\Omega-\ce_e))\cosh(\pi(\Omega+\omega-\ce_e))}\\
  &\times\left(V_h^\mathrm{A} \mathcal{R}_{h}^{\textrm{A}}\big(\Omega-\calE_{e}\big)+V_h^\mathrm{S}\mathcal{R}_{h}^{\textrm{S}}\big( \Omega-\calE_{e}\big)+V_h^\mathrm{A} \mathcal{R}_{h}^{\textrm{A}}\big(\Omega+\omega-\calE_{e}\big)+V_h^\mathrm{S}\mathcal{R}_{h}^{\textrm{S}}\big( \Omega+\omega-\calE_{e}\big)\right)\,.
\end{split}
\end{equation} Using explicit expression \eqref{eq:RAandRS} for $\mathcal{R}_{h}^{\textrm{A/S}}$, we need the following auxilary integral ($\epsilon\to 0^+$)
\begin{equation}\label{}
\begin{split}
  J_1(\omega)&=\int \rd \Omega\frac{e^{\epsilon\Omega}}{ \omega} \left(\frac{1}{e^{2\pi \Omega}+1}-\frac{1}{e^{2\pi (\Omega+\omega)}+1}\right)\frac{\cosh(\pi\Omega)\cosh(\pi(\Omega+\omega))}{\cosh(\pi(\Omega-\ce_e))\cosh(\pi(\Omega+\omega-\ce_e))}\\
  &\quad \times \pFq{3}{2}{h,1-h,\frac{1}{2}+i(\Omega-\ce_e)}{1, 1}{1}\\
  &= \pFq{3}{2}{h,1-h,1-i\omega}{1, 2}{1}\,,
\end{split}
\end{equation} which is evaluated by expanding the hypergeometric function series, doing the integral term by term with contour method as above, and then resumming.

Using $J_1$, we can express the correction as 
\begin{equation}
    \Re\delta_h \sigma(\omega)=-\sigma_0\frac{2\left(\frac{\pi}{2}\right)^h\Gamma(h)}{\sqrt{\pi}\sin\left(\frac{\pi h}{2}\right)\Gamma(h-1/2)}\left(V_h^\mathrm{A} \Re(J_1(\omega)+J_1(-\omega))+V_h^\mathrm{S}\Im(J_1(\omega)+J_1(-\omega))\right)\,.
\end{equation} Here the $J_1(-\omega)$ terms come from integrating $\mathcal{R}_{h}^{\textrm{A/S}}(\Omega+\omega-\ce_e)$. Since $J_1(\omega)=J_1(-\omega)^*$, only the $V_h^\mathrm{A}$ terms are nonzero, and we obtain \eqref{eq:sigmah=2}.

\bibliography{dqcp}
\end{document}